\documentclass[journal]{IEEEtran}
\usepackage{cite}
\usepackage{amsmath,amssymb,amsfonts}
\usepackage{algorithmic}
\usepackage{graphicx}
\usepackage{textcomp}
\usepackage{xcolor}
\usepackage{algorithm}

\usepackage{booktabs}       % professional-quality tables

\ifCLASSINFOpdf
\else
\fi
\begin{document}
\title{A Hybrid Residual Dilated LSTM end Exponential Smoothing Model for Mid-Term Electric Load Forecasting}

\author{Grzegorz~Dudek,
	Pawe\l~Pe\l ka,
	and~Slawek~Smyl% <-this % stops a space

\thanks{G. Dudek and P. Pe\l ka are with the Department 
	of Electrical Engineering, Czestochowa University of Technology, 42-200 Czestochowa, Al. Armii Krajowej 17, Poland, e-mail: dudek@el.pcz.czest.pl, p.pelka@el.pcz.czest.pl.}% <-this % stops a space
\thanks{S. Smyl is from Uber Technologies, 555 Market St, 94104, San Francisco, CA, USA, e-mail: slaweks@hotmail.co.uk.}
\thanks{Manuscript received ..., ...; revised ..., ....}}% <-this % stops a space	

\maketitle

\begin{abstract}
	
	This work presents a hybrid and hierarchical deep learning model for mid-term load forecasting. The model combines exponential smoothing (ETS), advanced Long Short-Term Memory (LSTM) and ensembling. ETS extracts dynamically the main components of each individual time series and enables the model to learn their representation. Multi-layer LSTM is equipped with dilated recurrent skip connections and a spatial shortcut path from lower layers to allow the model to better capture long-term seasonal relationships and ensure more efficient training. A common learning procedure for LSTM and ETS, with a penalized pinball loss, leads to simultaneous optimization of data representation and forecasting performance.
	In addition, ensembling at three levels ensures a powerful regularization. A simulation study performed on the monthly electricity demand time series for 35 European countries confirmed the high performance of the proposed model and its competitiveness with classical models such as ARIMA and ETS as well as state-of-the-art models based on machine learning. 	 
\end{abstract}


% \fi
\IEEEpeerreviewmaketitle

\section{Introduction}
\IEEEPARstart{E}{}lectricity demand forecasting is an essential tool in all sectors in the electric power industry. Mid-term load forecasting (MTLF), which involves forecasting the daily peak load for future months as well as monthly electricity demand, is necessary for power system operation and planning in such areas as maintenance scheduling, fuel reserve planning, hydro-thermal coordination, planing of electrical energy import and export and also security assessment. In deregulated power systems, MTLF is a basis for the negotiation of forward contracts. Forecast accuracy translates directly into financial performance for energy companies and energy market participants. The financial impact can be measured in millions of dollars for every point of forecasting accuracy gained. All the above reasons justify interest in new forecasting tools for MTLF.

In this work, we focus on monthly electricity demand forecasting. Monthly electricity demand time series express a nonlinear trend, yearly cycles and a random component. The trend is dependent on the rate of economic development in a country, while seasonal cycles are related to the climate, weather factors and the variability of seasons \cite{Apa12}. Factors disrupting electricity demand in a mid-term horizon include unpredictable economic events and political decisions \cite{Dog16}. 

MTLF methods can be divided into statistical/econometric models or machine learning/computational intelligence models \cite{Sug11}. Typical examples of the former are ARIMA, EST, and linear regression. ARIMA and ETS can deal with seasonal time series but linear regression requires additional operations such as decomposition or the extension of the model with periodic components \cite{Bar01}. Statistical models have undoubted advantages, being relatively simple, robust, efficient, and automatic, so that they
can be used by non-expert users.

The flexibility of machine learning (ML) models has increased researchers' interest in them as MTLF tools \cite{Gon08}. Of these, neural networks (NNs) are the most explored because of their attractive features such as learning capability, universal approximation property, nonlinear modeling, massive parallelism and ease of specifying a loss function, to align it better with forecasting goals. Some examples of using different architectures of NNs for MLTF are: \cite{Chen17} where NN learns on historical demand and weather factors, \cite{Gav01} where Kohonen NN was used, \cite{Dov99} where NNs were supported by fuzzy logic, \cite{Pel19} where generalized regression NN was used, \cite{Ahm19} where NNs, linear regression and AdaBoost were used, \cite{Pei11} where weighted evolving fuzzy NNs were combined, and \cite{Bae19} where recurrent NNs were used. Among other machine learning MTLF models, the following can be mentioned: support vector machines \cite{Zhao12}, and pattern similarity-based models \cite{Dud20}.

Recent trends in ML such as deep learning, especially deep recurrent NNs (RNNs), are very attractive for time series forecasting \cite{Hew19}. RNNs with connections between nodes forming a directed graph along a temporal sequence are able to exhibit temporal dynamic behavior using their internal state (memory) to process sequences of inputs. Recent works have reported that RNNs, such as the LSTM, provide high accuracy in forecasting and outperform most of the traditional statistical and ML methods such as ARIMA, support vector machine and shallow NNs \cite{Yan18}. There are many examples of an application of LSTMs to load forecasting: \cite{Bed18}, \cite{Zhe17}, \cite{Nar17}.

Many new ideas in the field of deep learning have been successfully applied to time series forecasting. For example in \cite{Tou19}, bidirectional LSTM is proposed for short-term scheduling in power markets. This solution has two benefits: long-range memory and bidirectional processing. It takes advantage of deep architectures, which are able to build up progressively higher level representations of data, by piling up RNN layers on top of each other. A residual recurrent highway network for  learning deep sequence predictions was proposed in \cite{Zia18}. It contains highways within the temporal structure of the network for unimpeded information propagation, thus alleviating gradient vanishing problem. Hierarchical structure learning is posed as a residual learning framework to prevent performance degradation problems. Another example of using a new deep learning solution for time series forecasting is the N-BEATS model proposed is \cite{Ore20}. Its architecture is based on backward and forward residual links and a deep stack of fully-connected layers. N-BEATS has a number of desirable properties, being interpretable, applicable without modification to a wide array of target domains, and fast to train. 

In addition to point forecasting, deep learning also enables probabilistic forecasting. A model proposed in\cite{Sal19}, DeepAR, produces accurate probabilistic forecasts, based on
training an auto-regressive RNN on a large number of related time series. This model makes probabilistic forecasts in the form of Monte Carlo samples that can be used to compute consistent quantile estimates for all sub-ranges in the prediction horizon. Another solution for probabilistic time series forecasting was proposed in \cite{Ran18}. It combines state space models with deep learning. By parametrizing a per-time-series linear state space model with a jointly-learned RNN, the method retains the desired properties of state space models such as data efficiency and interpretability, while making use of the ability to learn complex patterns from raw data offered by deep learning approaches. 

To improve forecasting performance, RNN is also mixed with other methods such as ETS. Such a model won the M4 forecasting competition in 2018 \cite{Mar18a}. This competition utilized 100,000 real-life time series, and incorporates all major forecasting methods, including those based on AI and ML, as well as traditional statistical ones \cite{Mar20}. The winning model, developed by Slawek Smyl \cite{Smy20}, is a hybrid approach utilizing both statistical and ML features. It combined ETS with advanced LSTM, which is supported by such mechanisms as dilation, residual connections and attention \cite{Cha17}, \cite{Kim17}, \cite{Qin17}. It produced the most accurate forecasts as well as the most precise prediction intervals. According to sMAPE, it was close to $10\%$ more accurate than the combination benchmark of the competition, which is a huge improvement. 
For monthly data (48,000 time series) it outperformed all other 60 submissions achieving the highest accuracy according to each of the three performance measures.

In this work, we propose a forecasting model for MTLF based on the winning submission to the M4 competition for monthly data. It combines ETS, LSTM and ensembling. ETS enables the model to capture the main components of the individual time series, such as seasonality and level, while LSTM allows nonlinear trends and cross-learning. A common learning procedure for LSTM and ETS, with a penalized pinball loss, leads to simultaneous optimization of data representation and forecasting performance. Ensembling at three levels is a powerful regularization method which reduces the model variance.  

The rest of the work is organized as follows. Section 2 describes the proposed forecasting model: its architecture, features, components and implementation details as well as data flow and processing. Section 3 describes the experimental framework used to evaluate the performance of the proposed model. Finally, Section 4 concludes the work.

%\section{Learning from M4 Competition}

\section{Forecasting Model}

The proposed model is based on the winning submission to the M4 forecasting competition 2018 for monthly data and point forecasts \cite{Smy20}. It is a hybrid and hierarchical forecasting model that enables ETS and advanced LSTM to be mixed into a common framework. The model architecture, its specific features and components are describe below.

\subsection{Framework and Features}

\begin{figure}[]
	\centering
	\includegraphics[width=0.4\textwidth]{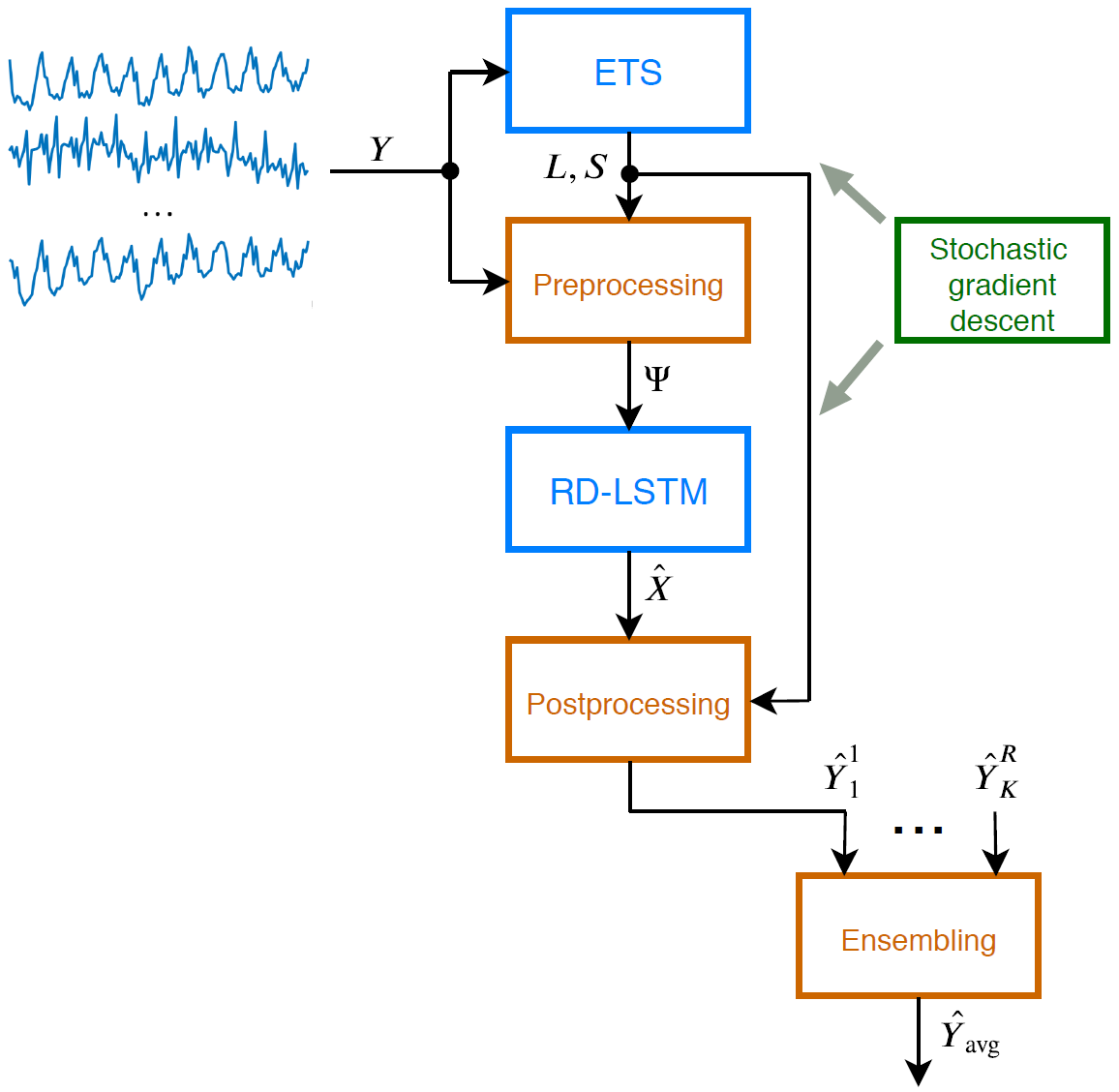}
	\caption{Block diagram of the ETS+RD-LSTM forecasting system.} 
	\label{figBd}
\end{figure}

The proposed forecasting model is shown in Fig. \ref{figBd}. It is composed of:
\begin{itemize}
	\item
	ETS -- which is a Holt-Winters type multiplicative seasonal model. It is used for extracting two components from the time series: level and seasonality. ETS loads a set of time series ($Y$), calculates the level and seasonal components individually for each series and returns sets of levels ($L$) and seasonal components ($S$).   
	
	\item
	Preprocessing -- the level and seasonal components are used for deseasonalization and adaptive normalization of the time series. The inputs to the preprocessing module are: set of time series $Y$ and sets of level and seasonal components, $L$ and $S$, respectively. The preprocessed data are divided into input and output training data and returned in training set $\Psi$.
	
	\item
	RD-LSTM -- which is residual dilated LSTM composed of four layers. Due to its recurrent nature, this model is capable of learning long-term dependencies in sequential data. RD-LSTM learns in cross-learning mode on training set $\Psi$. The forecasts for all time series produced by RD-LSTM are returned in set $\hat{X}$.
	
	\item
	Postprocessing -- the forecasts of the deseasonalized and normalized time series are "reseasonalised" and renormalised. The inputs to the postprocessing module are: forecasts $\hat{X}$, and level and seasonality sets, $L$ and $S$. The output is set $\hat{Y}$, containing the forecasts for each time series.
	
	\item
	Ensembling -- the forecasts produced by individual models are averaged. This enhances the robustness of the method further, mitigating model and parameter uncertainty. The ensembling module receives the sets of forecasts produced by individual models, $\hat{Y}_k^r$, aggregates them and returns a set of forecasts for all time series, $\hat{Y}_{\text{avg}}$.
	
	\item
	Stochastic gradient descent (SGD) -- the parameters of both ETS and RD-LSTM are updated by the same overall optimization procedure, SGD, with the overarching goal of minimizing forecasting errors.       
\end{itemize}	
	
The proposed model has a hierarchical structure, i.e. the data are exploited in a hierarchical manner. Both local and global time series features are extracted. The global features are learned by RD-LSTM across many time series. The specific features of each individual time series are extracted by ETS. Thus, each series has a partially unique and partially shared model.

Note the hybrid structure of the model, where statistical modeling is combined concurrently with ML algorithms. The model combines ETS, advanced LSTM and ensembling. ETS is focused on each individual series and enables the model to capture its main components such as seasonality and level. These components are used for time series preprocessing, normalization and deseasonalization.

An advanced LSTM-based RNN allows non-linear trends and cross-learning. This is an extended, multilayer version of LSTM with residual dilated LSTM blocks. The dilated recurrent skip connections and spatial shortcut path from lower layers, applied in this solution, allow the model to better capture long-term seasonal relationships and ensure more efficient training. The RD-LSTM model is trained on many time series (cross-learning). To train deep NNs, which have many parameters, cross-learning is necessary. Moreover, it enables the method to capture the shared features and components of the time series. 

ETS and RD-LSTM are optimized simultaneously, i.e. the ETS parameters and the RD-LSTM weights are optimised by SGD at the same time. The same overall learning procedure optimizes the model, including data preprocessing. So, the learning process includes representation learning -- searching for the most suitable representations of input and output data (individually for each time series), which ensures the most accurate forecasts. It is worth noting the dynamical character of the training set which is related to representation learning. The training set is updated in each epoch of RD-LSTM learning. This is because SGD updates the ETS parameters in each epoch, and therefore the level and seasonal components, used for preprocessing, are updated as well.

Ensembling is seen as a much more powerful regularization technique
than more popular alternatives, e.g. dropout or L2-norm penalty \cite{Ore20}. In our case, ensembling combines individual forecasts at three levels: stage of training level, data subset level and model level. This reduces the variance related to the stochastic nature of SGD, and also related to data and parameter uncertainty. 

\subsection{Exponential Smoothing}

The complex nature of a time series, e.g. nonstationarity, non-linear trend and seasonal variations, makes forecasting difficult and puts high demands on the models. A typical approach in this case is to simplify the forecasting problem by deseasonalization, detrending or decomposition. A time series is usually decomposed into seasonal, trend and stochastic components. The components expressing less complexity than the original time series can be modeled independently using simpler models. The most popular methods of decomposition are \cite{Hyn20}: additive decomposition, multiplicative decomposition, X11, SEAT, and STL.
Although very useful, this approach has a drawback. It separates the preprocessing from the forecasting, which results in the final solution not being optimal. Some classic statistical models, such as ETS, employ a better way: the forecasting model has a built-in mechanism to deal with seasonality. The final model uses the optimal decomposition of the time series. 

In our approach, we use ETS as the preprocessing tool. ETS extracts two components from the time series: level (smoothed value) and seasonality. Then we use these components to normalize and deseasonalize the original time series. Preprocessed time series are forecasted by RD-LSTM. ETS and RD-LSTM are optimised simultaneously using SGD. So the resulting forecasting model, including data preprocessing, is optimized as a whole. This distinctive feature of the proposed approach needs to be emphasized. 

The ETS model used in this study was inspired by the Holt-Winters multiplicative seasonal model. However, it has been simplified by the removal of the linear trend component. This is because the trend forecasting is the task of RD-LSTM, which is able to produce a non-linear trend which is more valuable in our case. The updating formulas for the ETS model with a seasonal cycle length of twelve (useful for monthly data) are as follows \cite{Smy20}:

\begin{equation}
\begin{aligned}
l_t=\alpha \frac{y_t}{s_t} + (1-\alpha)l_{t-1} \\
s_{t+12}=\beta \frac{y_t}{l_t} + (1-\beta)s_t
\label{eqls}
\end{aligned}
\end{equation} 
where $y_t$ is the time series value at timepoint $t$, $l_t$ and $s_t$ are the level and seasonal components, respectively, and $\alpha$, $\beta \in [0, 1]$ are smoothing coefficients.

\begin{figure}[]
	\centering
	\includegraphics[width=0.4\textwidth]{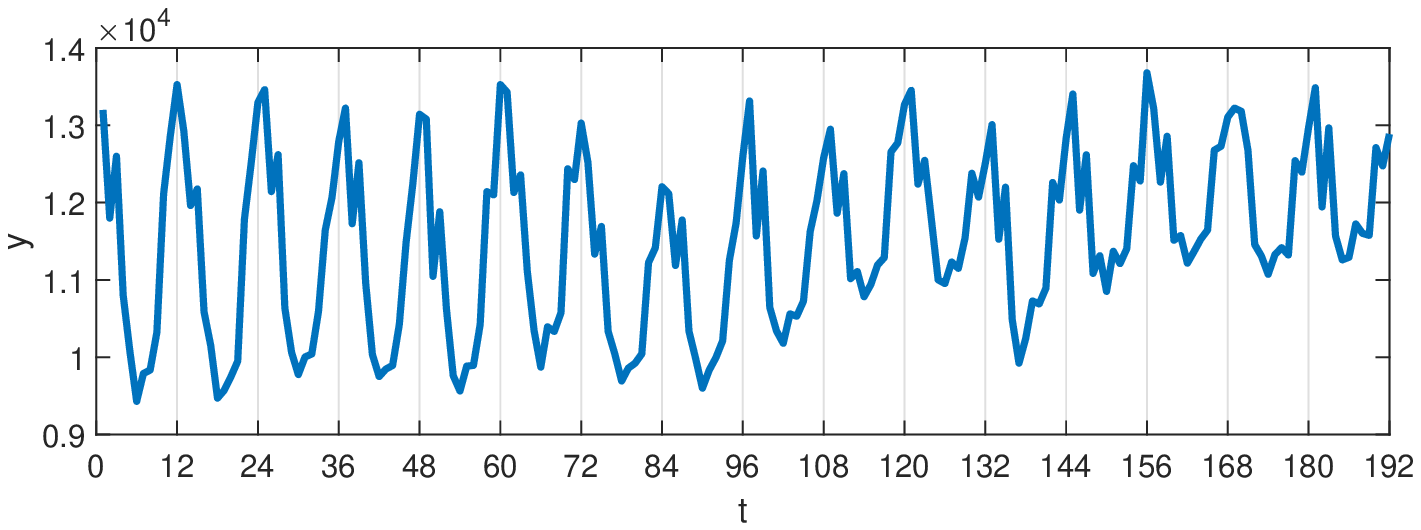}
	\includegraphics[width=0.4\textwidth]{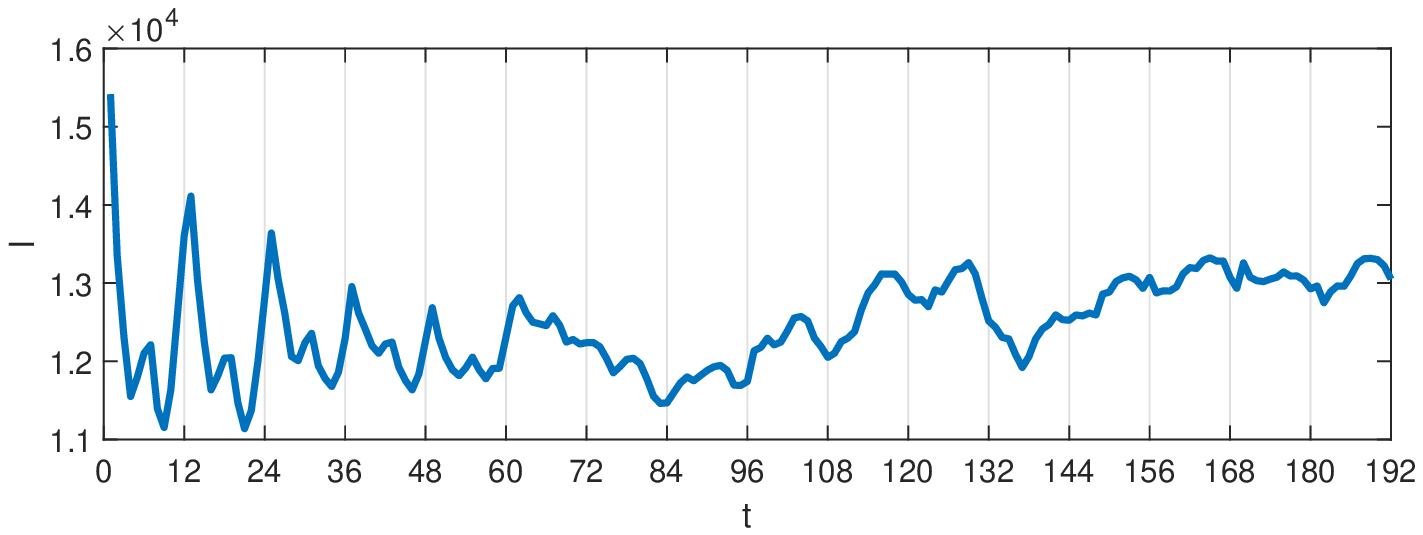}
	\includegraphics[width=0.4\textwidth]{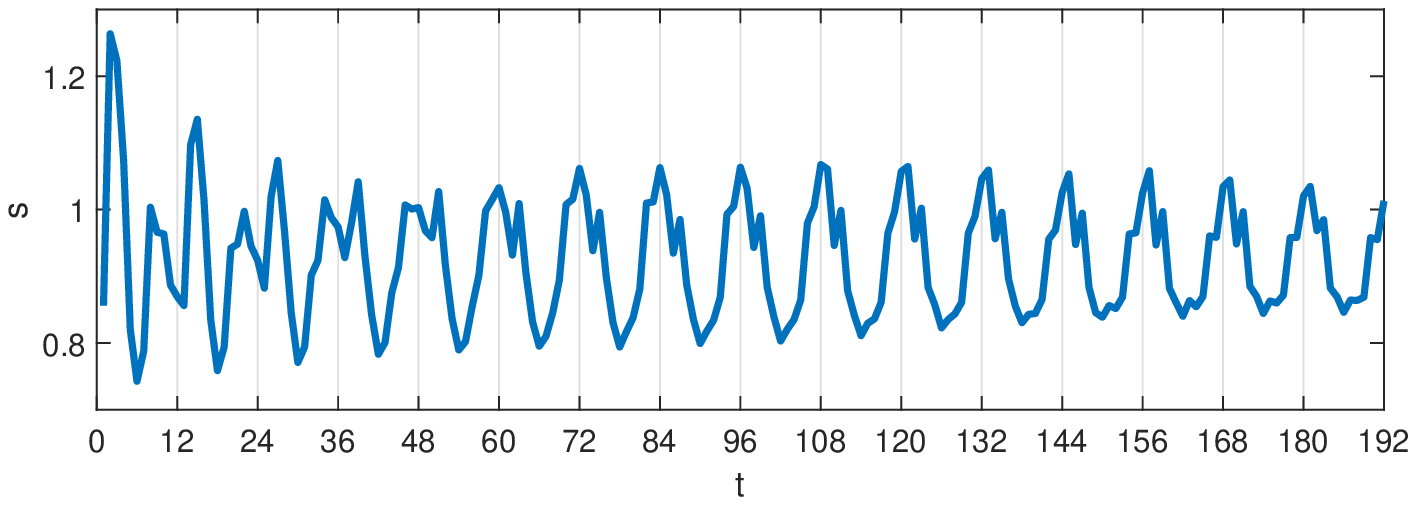}
	\caption{Original time series and its level and seasonal components.} 
	\label{figpls}
\end{figure}

The level equation shows a weighted average between the seasonally adjusted observation and the level for time $t-1$. The seasonal equation expresses a seasonal component for time $t+12$ as a weighted average between a new estimate of the seasonality component $(y_t/l_t)$ and the past estimate $(s_t)$. Fig. \ref{figpls} depicts an example of the monthly electricity demand time series and its level and seasonal components obtained from \eqref{eqls}.      

The ETS model parameters, twelve initial seasonal components and two smoothing coefficients for each time series, were adjusted together with RD-LSTM weights by SGD. Knowing these parameters allows the level and seasonal components to be calculated, which are then used for preprocessing: deseasonalization and normalization.

\subsection{Pre- and Postprocessing}

The level and seasonal components are calculated for all points of each series, which are then used for deseasonalization
and adaptive normalization during the on-the-fly preprocessing. This is the most crucial element of the forecasting procedure as it determines its performance. The time series is preprocessed in each training epoch using the updated values of level and seasonal components. These updated values are calculated from \eqref{eqls}, where the ETS parameters are increasingly fine tuned in each epoch by SGD. 

The time series is preprocessed using rolling windows: input and output ones. Both windows have a length of twelve, which is equal to the length of both the seasonal cycle and the forecast horizon. The input window, $\Delta_{in}$, contains twelve consecutive elements of the time series which after preprocessing will be the RD-LSTM inputs. The corresponding output window, $\Delta_{out}$, contains the next twelve consecutive elements, which after preprocessing will be the RD-LSTM outputs. The time series fragments inside both windows are normalized by dividing them by the last value of the level in the input window, $l_t^*$, and then, divided further by the relevant seasonal component. As a result of this operation, we obtain positive input and output values close to one. Finally, to limit the destructive impact of outliers on the forecasts, a squashing function, log(.), is applied. The resulting preprocessing can be expressed as follows:

\begin{equation}
x_t=\log \left(\frac{y_t}{l_t^* s_t} \right)
\label{eqxt}
\end{equation} 
where $x_t$ is the preprocessed $t$-th element of the time series, $l_t^*$ is the last value of the level in input window $\Delta_{in}$, and $s_t$ is the $t$-th seasonal component. 

Note that normalization is adaptive and local, and the "normalizer" follows the series values. This allows us to include the current features of the series ($l_t^*$ and $s_t$) in the input and output variables. 

The preprocessed elements of the time series contained in the successive input and output windows can be represented by vectors as follows: \\
-- first pair of input and output windows: \\ 
$\textbf{x}_1^{in} = [x_1 x_2 ... x_{12}]$, $\textbf{x}_1^{out} = [x_{13} x_{14} ... x_{24}]$, \\
-- second pair of input and output windows: \\ 
$\textbf{x}_2^{in} = [x_2 x_3 ... x_{13}]$, $\textbf{x}_2^{out} = [x_{14} x_{15} ... x_{25}]$, \\
-- \ldots, \\
-- $N$-th pair of input and output windows: \\ 
$\textbf{x}_N^{in} = [x_N x_{N+1} ... x_{N+11}]$, $\textbf{x}_N^{out} = [x_{N+12} x_{N+13} ... x_{N+23}]$.

These vectors are included in the training subset for the $i$-th time series: $\Phi_i=\{(\textbf{x}_t^{in}, \textbf{x}_t^{out}): t = 1, 2, ..., N\}$. The training subsets for all $M$ time series are combined and form the training set $\Psi=\{\Phi_1, \Phi_2, ..., \Phi_M\}$ which is used for RD-LSTM cross-learning. Note the dynamic character of the training set. It is updated in each epoch because the level and seasonal components in \eqref{eqxt} are updated.

\begin{figure}[]
	\centering
	\includegraphics[width=0.24\textwidth]{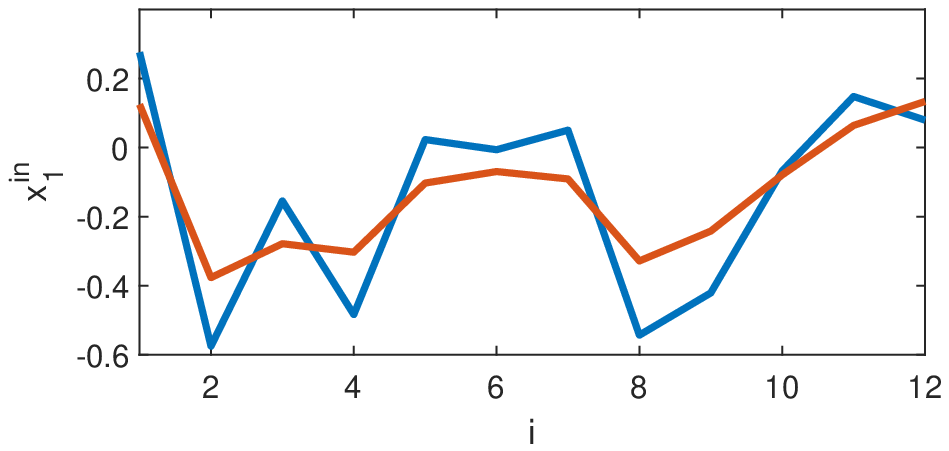}
	\includegraphics[width=0.24\textwidth]{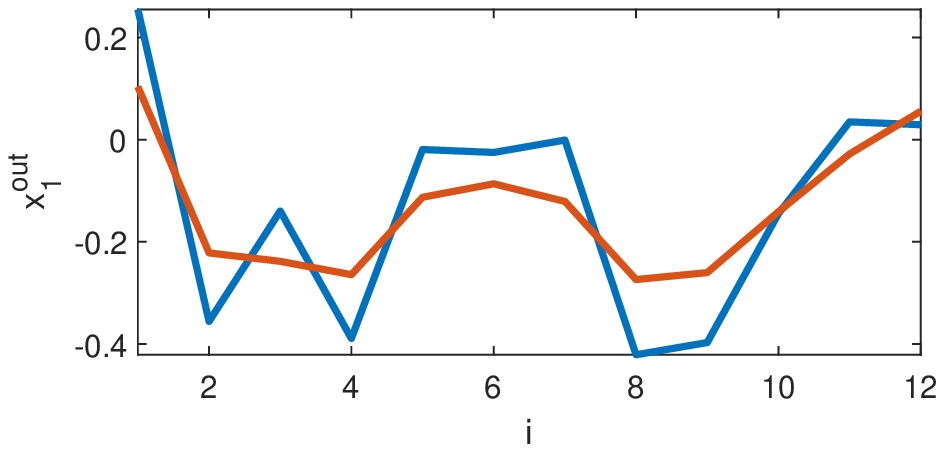}
	\includegraphics[width=0.24\textwidth]{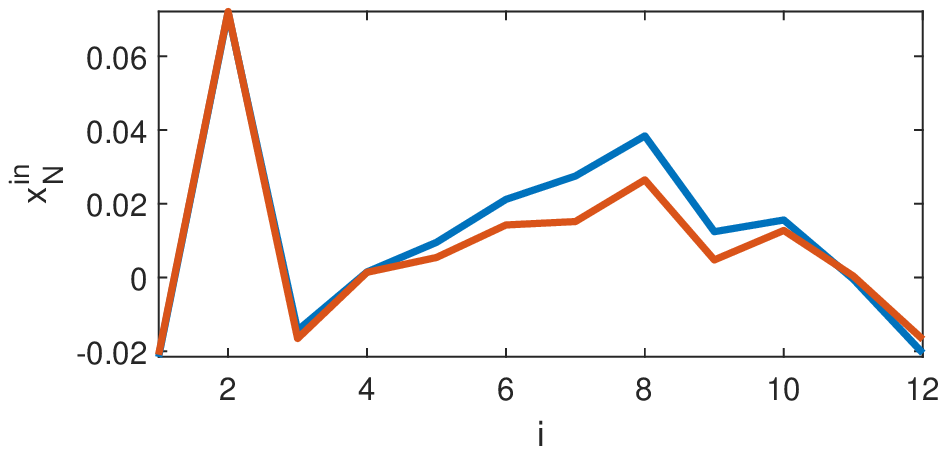}
	\includegraphics[width=0.24\textwidth]{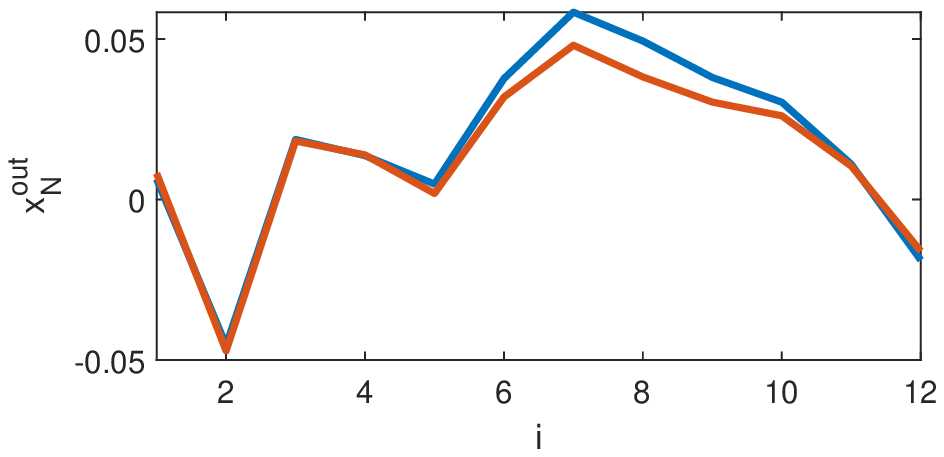}		
	\caption{Examples of the input (left panel) and output (right panel) vectors created for the time series shown in Fig. \ref{figpls}. Upper panel shows x-vectors for the first two cycles, lower panel shows x-vectors for the last two cycles. The x-vectors created in the first training epoch in blue, in the last epoch in red.}
	\label{figx}
\end{figure}

Fig. \ref{figx} shows the input and output vectors for the time series which is shown in Fig. \ref{figpls}. The upper panel shows the corresponding input and output x-vectors representing the first two seasonal cycles of the time series. The lower panel shows the input and output x-vectors representing the last two seasonal cycles. It can be seen from this figure that the x-vectors express patterns of the time series fragments after filtering out both level and seasonality. This pattern representation of the time series has been used successfully in earlier studies concerning ML forecasting models, especially similarity based models \cite{Dud15a}. Different definitions of the time series patterns can be found in \cite{Dud15}. But these definitions are fixed, while in this work we use dynamic patterns which change during learning (compare the patterns in the first and last training epochs in Fig. \ref{figx}).  
        
RD-LSTM operates on preprocessed time series values, $x_t$. In the postprocessing step, the forecasts generated by RD-LSTM, $\hat{x}_t$, need to be "unwound" in the following way:

\begin{equation}
\hat{y}_t=\exp (\hat{x}_t) l_t^* s_t
\label{eqyt}
\end{equation}

Note that both level $l_t^*$ and seasonal component $s_t$ in \eqref{eqyt}, which are necessary to calculate $\hat{y}_t$ from $\hat{x}_t$, are known. They are determined from \eqref{eqls} on the basis of the time series history. 

\subsection{Residual Dilated LSTM}

\begin{figure}[]
	\centering
	\includegraphics[width=0.3\textwidth]{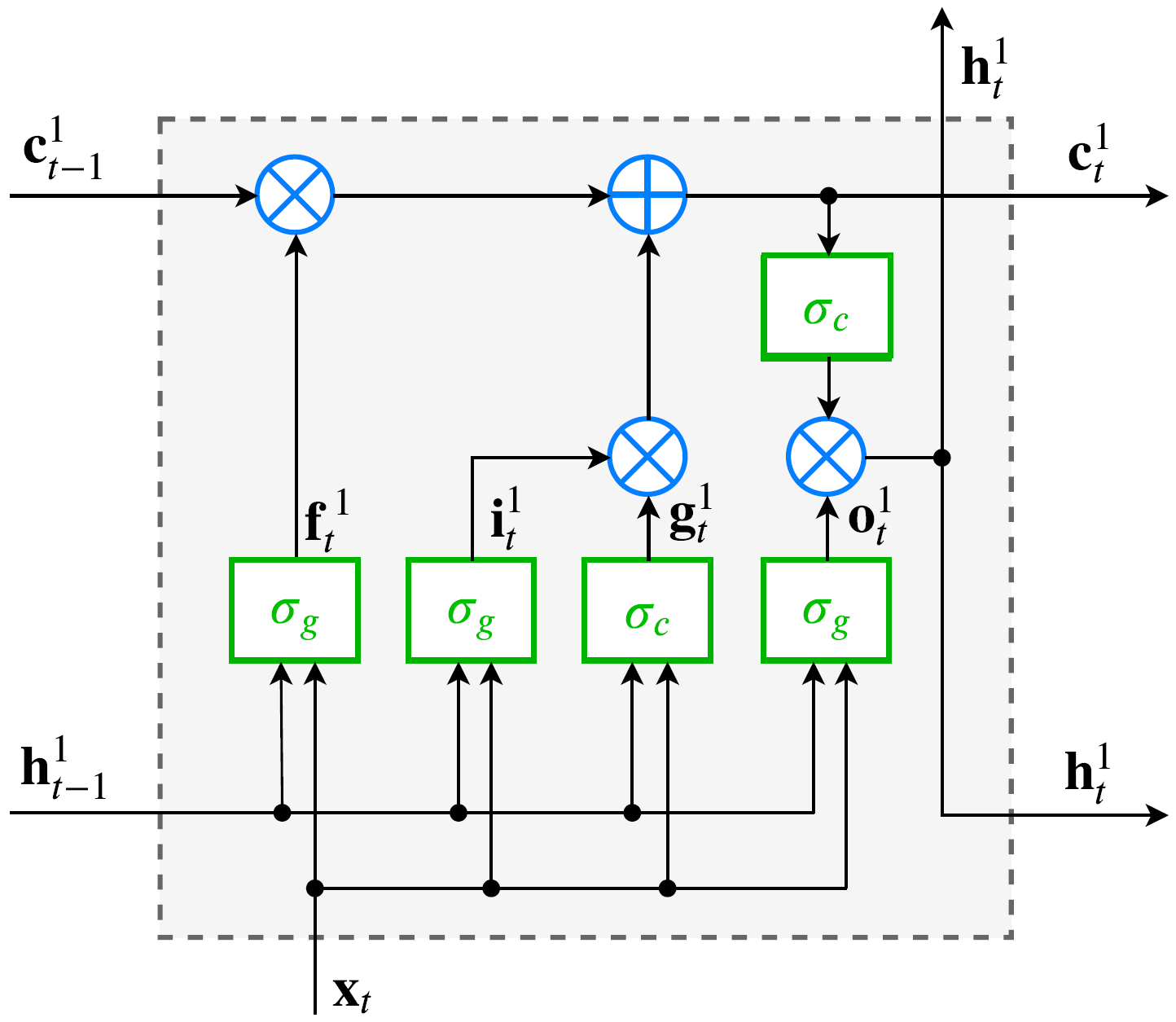}
	\caption{LSTM block.} 
	\label{figCe1}
\end{figure}

LSTM is a special kind of RNN capable of learning long-term dependencies in sequential data \cite{Hoch97}. A common LSTM block is composed of a memory cell which can maintain its state over time, and three non-linear "regulators", called gates, which control the flow of information inside the block. A typical LSTM block is shown in Fig. \ref{figCe1}. In this diagram, $\textbf{h}_t$ and $\textbf{c}_t$ denote the hidden state and the cell state at time step $t$, respectively. The cell state contains information learned from the previous time steps. Information can be added to or removed from the cell state using the gates: input gate ($i$), forget gate ($f$) and output gate ($o$). At each time step $t$, the block uses the past state of the network, i.e. $\textbf{c}_{t-1}$ and $\textbf{h}_{t-1}$, and the input $\textbf{x}_t$ to compute output $\textbf{h}_{t}$ and updated cell state $\textbf{c}_{t}$. The hidden and cell states are recurrently connected back to the block input. All of the gates are controlled by the hidden state of the past cycle and the input x-vector. Most modern studies incorporate many of the improvements that have been made to the LSTM architecture since its original formulation \cite{Gre17}.

Detailed mathematical expressions describing LSTM block are given below. The cell state at time step $t$ is:
\begin{equation}
\textbf{c}_t^1 = \textbf{f}_t^1  \otimes  \textbf{c}_{t-1}^1  + \textbf{i}_t^1 \otimes \textbf{g}_t^1
\label{eqL1}
\end{equation} 
where operator $\otimes$ denotes the Hadamard product (element-wise product) and superscript, 1, refers to the first layer of RD-LSTM network, where we use the standard LSTM block.

The hidden state at time step $t$ is given by:

\begin{equation}
\textbf{h}_t^1 = \textbf{o}_t^1   \otimes  \sigma_c(\textbf{c}_t^1) 
\label{eqL2}
\end{equation} 
where the state activation function $\sigma_c$ is a hyperbolic tangent function.

Formulas related to the gates are as follows:

\begin{equation}
\textbf{f}_t^1= \sigma_g(\textbf{W}_f^1 \textbf{x}_t+\textbf{V}_f^1\textbf{h}_{t-1}^1+ \textbf{b}_f^1)
\label{eqL3}
\end{equation} 
\begin{equation}
\textbf{i}_t^1= \sigma_g(\textbf{W}_i^1\textbf{x}_t+\textbf{V}_i^1\textbf{h}_{t-1}^1+ \textbf{b}_i^1)
\label{eqL4}
\end{equation} 
\begin{equation}
\textbf{g}_t^1= \sigma_c(\textbf{W}_g^1\textbf{x}_t+\textbf{V}_g^l\textbf{h}_{t-1}^1+\textbf{b}_g^1)
\label{eqL5}
\end{equation} 
\begin{equation}
\textbf{o}_t^1= \sigma_g(\textbf{W}_o^1\textbf{x}_t+\textbf{V}_o^l\textbf{h}_{t-1}^1+\textbf{b}_o^1)
\label{eqL6}
\end{equation} 
where $\textbf{W}$, $\textbf{V}$ and $\textbf{b}$ are input weights, recurrent weights and biases, respectively, and $\sigma_g$ is a gate sigmoid activation function  $(1+e^{-x})^{-1}$.

In our study, we also use a dilated residual version of LSTM (RD-LSTM). Dilated RNN architecture was proposed in \cite{Cha17} as a solution to tackle three major challenges of RNN when learning on long sequences: complex dependencies, vanishing and exploding gradients, and efficient parallelization. It is characterized by multi-resolution dilated recurrent skip connections. Moreover, it reduces the number of parameters needed and enhances training efficiency significantly in tasks involving very long-term dependencies. A dilated LSTM block receives as input states, not the last ones, $\textbf{c}_{t-1}$ and $\textbf{h}_{t-1}$, but earlier states, $\textbf{c}_{t-d}$ and $\textbf{h}_{t-d}$, where $d>1$ is a dilation. So, to compute the current states of the LSTM block, the last $d-1$ states are skipped. Usually multiple dilated recurrent layers are stacked with hierarchical dilations to construct a system, which learns the temporal dependencies of different scales at different layers. In \cite{Cha17}, it was shown that this solution can reliably improve the ability of recurrent models to learn long-term dependency in problems from different domains. It seems that dilated LSTM can be particularly useful for seasonal time series, where the relationships between the series elements have a cyclical character. This character can be incorporated into the model by dilations related to seasonality.

A residual version of LSTM was proposed in \cite{Kim17}. A standard memory cell, which learns long-term dependencies of sequential data, provides a temporal shortcut path to
avoid vanishing or exploding gradients in the temporal domain. The residual LSTM provides an additional spatial shortcut path from lower layers for efficient training of deep LSTM architectures. To avoid a conflict between spatial and temporal-domain gradient
flows, residual LSTM separates the spatial shortcut path from the temporal one. This gives greater flexibility to deal with vanishing or exploding gradients. 

\begin{figure}[]
	\centering
	\includegraphics[width=0.35\textwidth]{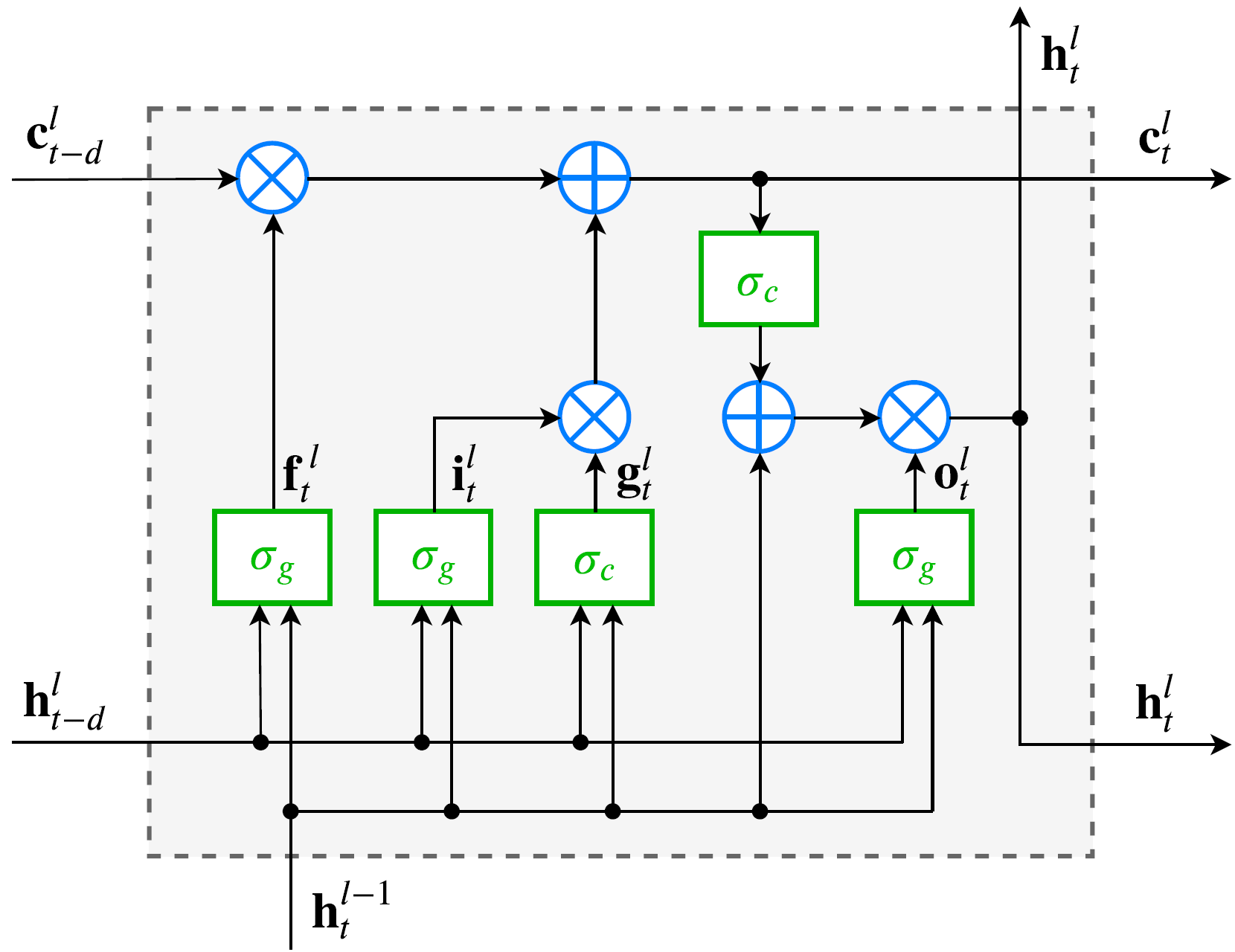}
	\caption{RD-LSTM block.} 
	\label{figCe2}
\end{figure}

Fig. \ref{figCe2} describes a residual dilated LSTM block which was used in this study. We denote this block by RD-LSTM. In this figure $\textbf{h}_t^{l-1}$ is a shortcut path from $(l-1)$-th layer that is added to updated cell state $\textbf{c}_t$ processed by function $\sigma_c$. Our implementation of the residual LSTM is a simplified version of the original one. The peephole connections are removed as well as linear transformations of the shortcut path $\textbf{h}_t^{l-1}$ and transformed cell state $\sigma_c(\textbf{c}_t^l)$. These transformations are not necessary because in our case the dimensions of $\textbf{h}_t^{l-1}$ and $\sigma_c(\textbf{c}_t^l)$ match that of $\textbf{h}_t^{l}$. 

The mathematical expressions describing the RD-LSTM block are as follows:

\begin{equation}
\textbf{c}_t^l = \textbf{f}_t^l  \otimes  \textbf{c}_{t-d}^l  + \textbf{i}_t^l \otimes \textbf{g}_t^l
\label{eqL1a}
\end{equation} 
\begin{equation}
\textbf{h}_t^l = \textbf{o}_t^l   \otimes  (\sigma_c(\textbf{c}_t^l) + \textbf{h}_t^{l-1}) 
\label{eqL2a}
\end{equation} 
\begin{equation}
\textbf{f}_t^l= \sigma_g(\textbf{W}_f^l \textbf{h}_t^{l-1}+\textbf{V}_f^l\textbf{h}_{t-d}^l+ \textbf{b}_f^l)
\label{eqL3a}
\end{equation} 
\begin{equation}
\textbf{i}_t^l= \sigma_g(\textbf{W}_i^l\textbf{h}_t^{l-1}+\textbf{V}_i^l\textbf{h}_{t-d}^l+ \textbf{b}_i^l)
\label{eqL4a}
\end{equation} 
\begin{equation}
\textbf{g}_t^l= \sigma_c(\textbf{W}_g^l\textbf{h}_t^{l-1}+\textbf{V}_g^l\textbf{h}_{t-d}^l+\textbf{b}_g^l)
\label{eqL5a}
\end{equation} 
\begin{equation}
\textbf{o}_t^l= \sigma_g(\textbf{W}_o^l\textbf{h}_t^{l-1}+\textbf{V}_o^l\textbf{h}_{t-d}^l+\textbf{b}_o^l)
\label{eqL6a}
\end{equation} 
where superscript $l$ indicates the layer number (from 2 to 4 in our case, see below) and $d$ is a dilation (3, 6 or 12 in our case, see below).

\begin{figure}[]
	\centering 
	\includegraphics[width=0.20\textwidth]{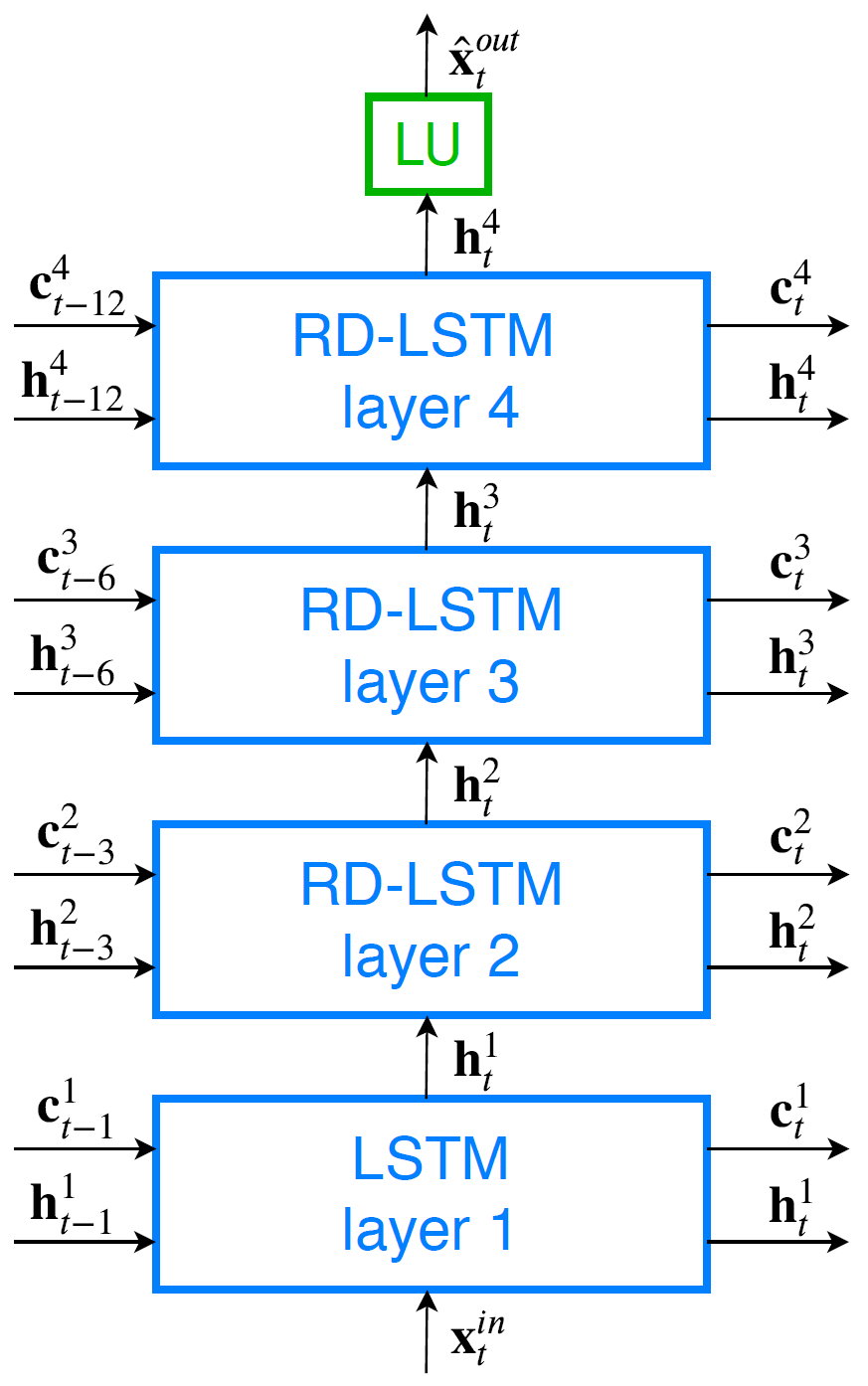}
	\caption{RD-LSTM network architecture.} 
	\label{figA}
\end{figure}

The proposed RD-LSTM architecture, which is a result of painstaking experimentation on M4 monthly data (48,000 time series), is depicted in Fig. \ref{figA}. It is composed of four recurrent layers and a linear unit LU. The first layer consists of the standard LSTM block shown in Fig. \ref{figCe1}. The subsequent three layers consist of RD-LSTM blocks (Fig. \ref{figCe2}) with increasing dilations $d=3, 6$ and $12$. The last element is a linear unit which transforms the output of the last layer, $\textbf{h}_t^4$, into the forecast of the output x-vector:

\begin{equation}
\hat{\textbf{x}}_t^{out} = \textbf{W}_x\textbf{h}_{t}^4+\textbf{b}_x   
\label{eqL7}
\end{equation} 

The learnable parameters of RD-LSTM are: input weights $\textbf{W}$, recurrent weights $\textbf{V}$, and biases $\textbf{b}$. They are tuned in the cross-learning mode simultaneously with the ETS parameters using SGD. The length of the cell and hidden states, $m$, the same for all layers, was selected on the training set (see Section III for details).  

Note that an input is not a scalar but a vector representing a sequence of the time series of length 12, i.e. a seasonal period. It allows the RD-LSTM to be exposed to the immediate history of the series directly. An output is a vector representing the whole forecasted sequence of length 12. 

When the output x-pattern is determined from \eqref{eqL7}, the forecasted monthly demands are calculated from \eqref{eqyt}. In the latter equation $\hat{x}_t$ is the $t$-th component of vector $\hat{\textbf{x}}_{\text{avg}}^{out}$.

\subsection{Ensembling}

The forecasts generated by the RD-LSTM model are ensembled at three levels:
\begin{enumerate}
	\item Stage of training level. Averaging forecasts produced by $L$ most recent training epochs. 
	\item Data subset level. Averaging forecasts from a pool of $K$ forecasting models which learns on the subsets of the training set. 
	\item Model level. Averaging forecasts from $R$ independent runs of the pool of $K$ models produced on the data subset level. 
   
\end{enumerate} 

%https://machinelearningmastery.com/polyak-neural-network-model-weight-ensemble/
The idea behind using the averaging forecasts produced in a few most recent training epochs (first level of ensembling) is that averaging has the effect of calming down the noisy SGD optimization process. SGD uses mini-batches of training samples to estimate the actual gradients. The resulting searching process operates on the approximated gradients which make the trajectory noisy. Averaging the forecasts obtained in the most recent epochs, when the algorithm converges around the local minimum, may reduce the effect of stochastic searching and produce more accurate forecasts.    

%At the data subset level, the forecasts from $K$ models which learns on the subsets of the training set, $\Psi$-subsets, are averaged. The training subsets $\Psi$ are created similarly to the training subsets in the cross-validation procedure. To get the $\Psi$-subsets, the training set composed of the subsets for all time series, $\Psi=\{\Phi_1, \Phi_2, ..., \Phi_M\}$, is randomly partitioned into $K$ equal sized subsets (each of them contains different $\Phi$-subsets). Of the $K$ subsets, a single subset is removed, and the remaining $K-1$ subsets form the $\Psi$-subsets. This process is then repeated $K$ times, with each of the $K$ subset removed exactly once. 

At the data subset level, the forecasts from $K$ models which learn on the subsets of the training set, $\Psi_1$, $\Psi_2$, ..., $\Psi_K$, are averaged. The training set, $\Psi=\{\Phi_1, \Phi_2, ..., \Phi_M\}$, is composed of subsets $\Phi_i$ containing the training samples for the $i$-th time series. To create the training $\Psi$-subsets, first, a set of $M$ time series is split randomly into $K$ subsets of similar size: $\Theta_1$, $\Theta_2$, ..., $\Theta_K$. The $k$-th $\Psi$-subset contains the  $\Phi$-subsets for all time series excluding those in $\Theta_k$, i.e. $\Psi_k=\Psi \backslash \{\Phi_i\}_{i \in \Theta_k}$. Each of $K$ models learns on its own training subset, $\Psi_k$, and generates forecasts for the time series included in $\Psi_k$. Then the $K-1$ forecasts for each time series produced by the pool of $K$ models are averaged.

The last level of ensembling simply averages the forecasts for each time series generated in $R$ independent runs of a pool of $K$ models. In each run, the training subsets $\Psi_k$ are created anew. 

Note that the diversity of learners, which is a key property which governs the ensemble performance \cite{Pet18}, has various sources in our proposed approach. They include (i) data uncertainty: learning on mini-batches, learning on different $\Psi$-subset of the training set, and (ii) parameter uncertainty: learning using different initial values of the model parameters in each run.

\begin{figure}[]
 	\centering
 	\includegraphics[width=0.40\textwidth]{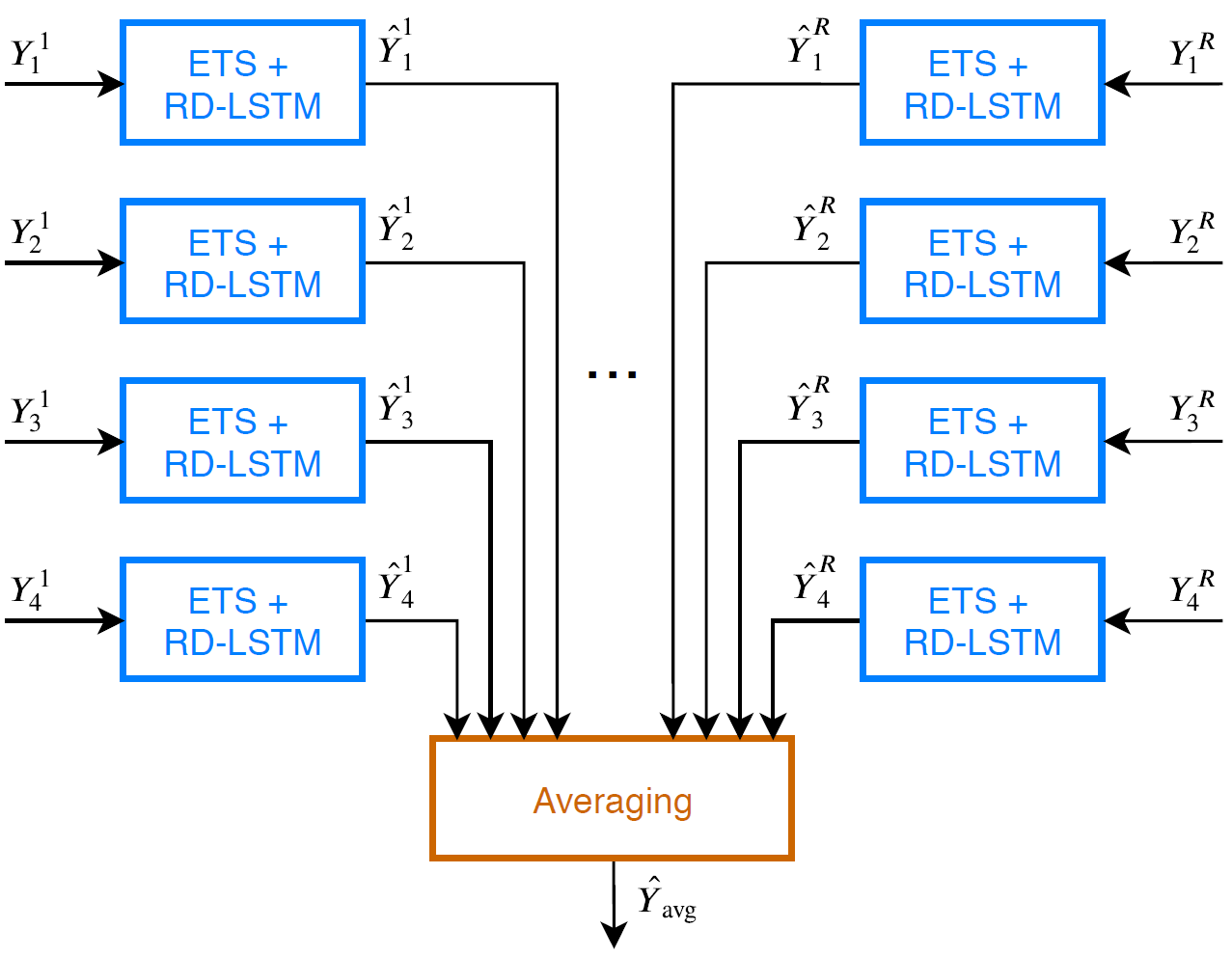}
 	\caption{Ensembling.} 
 	\label{figE}
 \end{figure}

The last two ensembling levels are depicted in Fig. \ref{figE}. In this figure, $K=4$, so four training $\Psi$-subsets are created. The set of time series included in the $k$-th $\Psi$-subsets in the $r$-th run is denoted by $Y_k^r$ in this figure, and the set of forecasts generated by the model in this case is denoted by $\hat{Y}_k^r$. For each time series $R(K-1)$ forecasts are averaged. This is shown as a joint operation for levels 2 and 3 in the figure, and can be expressed as:

  \begin{equation}
  \hat{\textbf{y}}_{\text{avg}} = \frac{1}{R(K-1)}\sum_{r=1}^{R}\sum_{k=1}^{K-1} \hat{\textbf{y}}_{r,k}
  \label{eqL8}
  \end{equation} 
  where $K$ is the size of the pool of models, $R$ is the number of runs, and $\hat{\textbf{y}}_{r,k}$ is the forecasted y-vector generated as the $k$-th one in the $r$-th run.
 
In this work, we use a simple average for ensembling, but other functions, such as median, mode, or trimmed mean could be applied \cite{Pet18}. As shown in \cite{Cha18}, a simple average of forecasts often outperforms more complicated weighting schemes.

Note that the $K \cdot R$ models created at the data subset levels in $R$ runs can be trained simultaneously. Thus, the proposed forecasting system is suitable to be implemented in parallel.

\subsection{Loss Function}

As a loss function we used pinball loss operating on normalized and deseasonalized RD-LSTM outputs and actuals:

\begin{equation}
L_t =
\begin{cases}
(x_t-\hat{x}_t)\tau       & \text{if } x_t \geq \hat{x}_t\\
(\hat{x}_t-x_t)(1-\tau)  &\text{if } \hat{x}_t > x_t 
\end{cases}
\label{eqLt}
\end{equation}
where $\tau \in (0, 1)$ controls the function asymmetry.

When $\tau=0.5$ the loss function is symmetrical and penalizes positive and negative deviations equally. When the model tends to have a positive or negative bias, we can reduce the bias by introducing $\tau$ smaller or larger than $0.5$, respectively. Thus, the asymmetric pinball function, penalizing positive and negative deviations differently, allows the method to deal with bias. It is worth noting that pinball loss is commonly employed in quantile regression and probabilistic forecasting \cite{Tak06}.

The experience gained during the M4 competition shows that the smoothness of the level time series influences the forecasting accuracy substantially. It turned out that when the input to the RD-LSTM was smooth, the RD-LSTM focused on predicting the trend, instead of overfitting on some spurious, seasonality-related patterns. A smooth level curve also means that the seasonal components absorbed the seasonality properly. To deal with the wiggliness of a level curve, a penalized loss function was introduced as follows:

\begin{itemize}
	\item Calculate the logarithms of the quotients of the neighboring points of the level time series: $d_t=\log(l_{t+1}/l_t)$,
	\item Calculate differences of the above: $e_t=d_{t+1}-d_t$,
	\item Square and average them for each series.   
\end{itemize}	
	
The resulting penalized loss function related to a given time series takes the form: 	
\begin{equation}
L = \frac{1}{T}\sum_{t=1}^{T} L_t + \lambda \frac{2}{T-2} \sum_{t=1}^{T-2} \log \left(
\frac{l_{t+2}l_t}{l_{t+1}^2} \right)
\label{eqL}
\end{equation}	
where $T$ is the number of forecasts and $\lambda$ is a regularization parameter which determines how much to penalizes the wiggliness of a level curve. 	 
	
The level wiggliness penalty affected the performance of the method significantly
and contributed greatly to winning the M4 competition \cite{Smy20}.

\section{Experimental Study}

\begin{figure}[]
	\centering
	\includegraphics[width=0.24\textwidth]{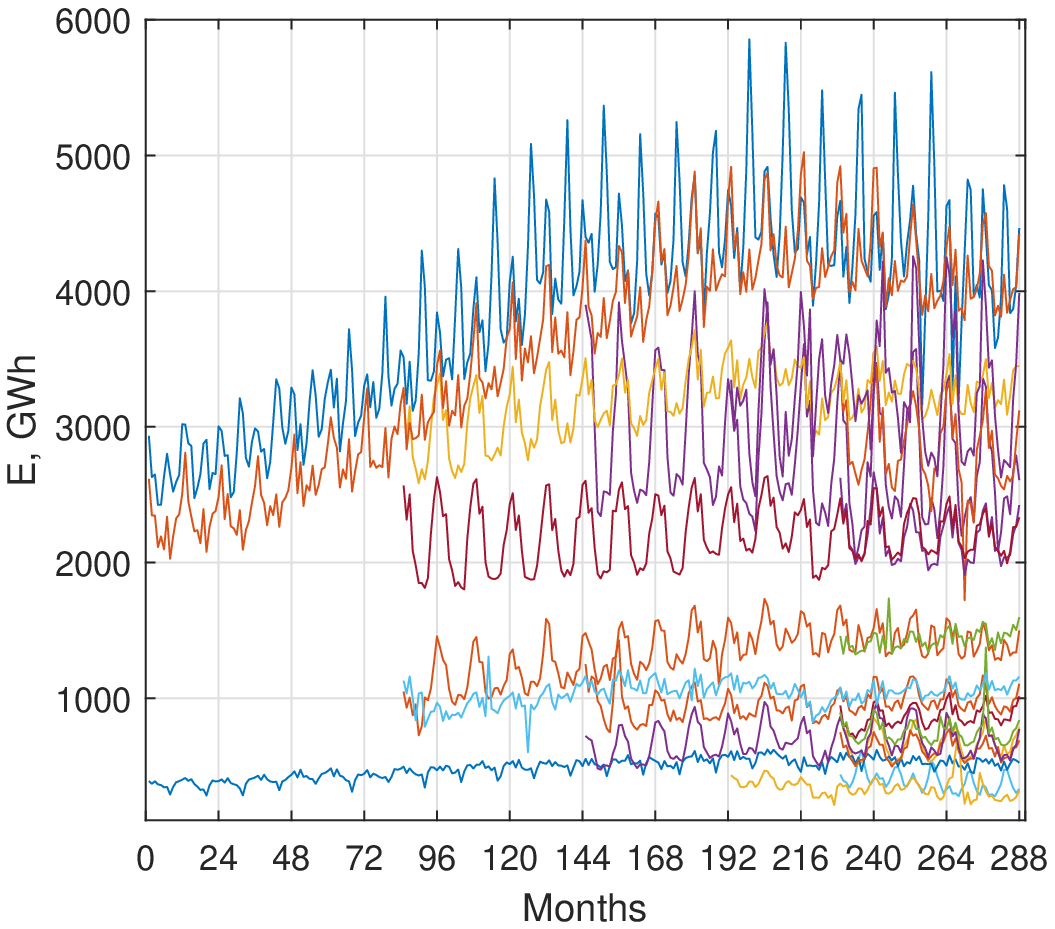}
	\includegraphics[width=0.24\textwidth]{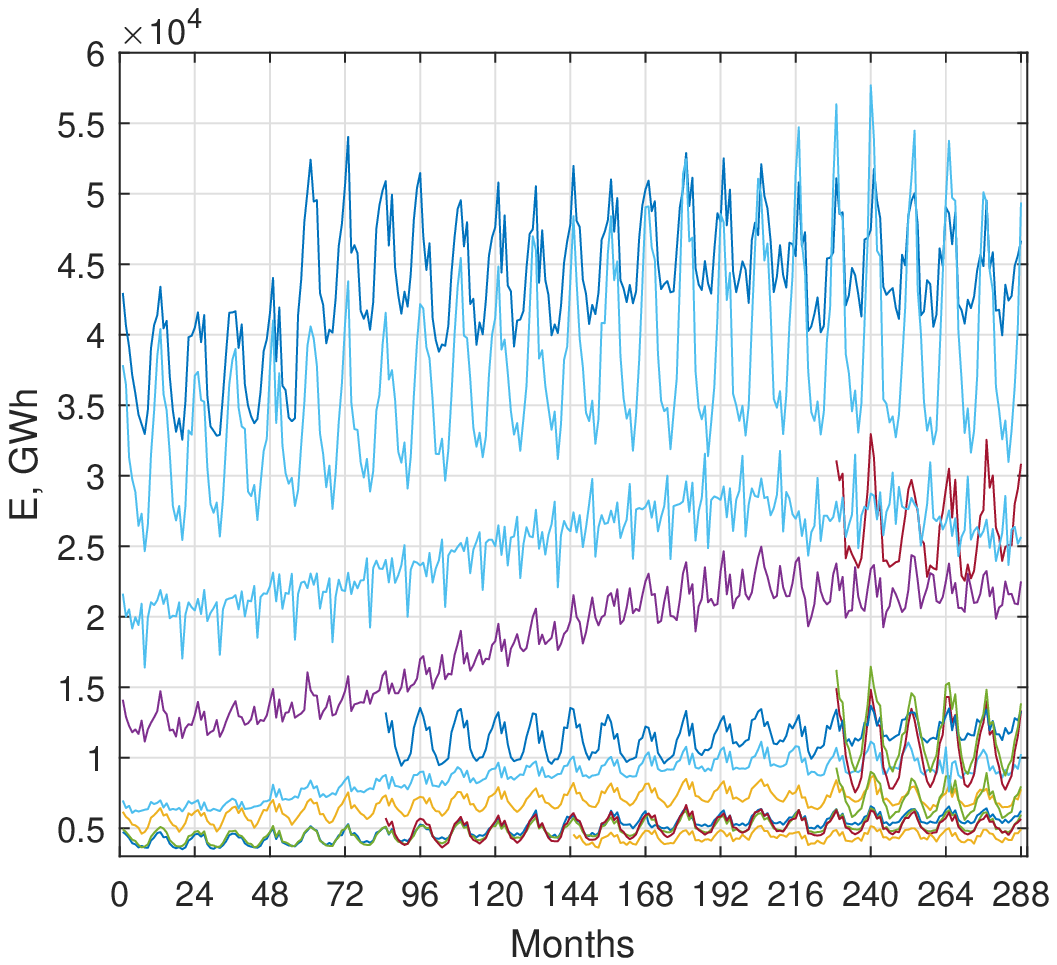}
	\caption{Monthly electricity demand time series for 35 European countries.} 
	\label{figTS}
\end{figure}

This section presents the results of applying the proposed forecasting model to the monthly electricity demand forecasting for 35 European countries. The data were obtained from the ENTSO-E repository (www.entsoe.eu). The time series have different lengths: 24 years (11 countries), 17 years (6 countries), 12 years (4 countries), 8 years (2 countries), and 5 years (12 countries). The last year of data is 2014. The time series are presented in Fig. \ref{figTS}. As can be seen from this figure, monthly electricity demand time series exhibit different levels, nonlinear trends, strong annual cycles and variable variances. The shapes of yearly cycles change over time. 

The forecasting model as well as comparative models were applied to forecast twelve monthly demands for the last year of data, 2014. The data from previous years were used for hyperparameter selection and learning.  

Model hyperparameters were selected on the training data. A typical procedure was learning the model on the time series fragments up to 2012, and validating it on 2013 (during the M4 competition a very strong correlation between errors for the test period and last period of training data was observed). The selected hyperparameters were used to construct the model for 2014. The selected hyperparameters were as follows:
    
\begin{itemize}
	\item number of epochs: $10$,
	\item learning rate: $10^{-3}$,
	\item length of the cell and hidden states: $m=40$,
	\item asymmetry parameter in pinball loss: $\tau=0.4$,
	\item regularization parameter: $\lambda=50$,
	\item ensembling parameters: $L=5$, $K=4$, $R=3$.
\end{itemize}

	The model was implemented in C++ relying on the DyNet library \cite{Neu17}. It was compiled in Visual Studio 2017 (Windows 10) and run in parallel on an 8-core CPU (AMD Ryzen 7 1700, 3.0 GHz, 32 GB RAM). 

\subsection{Comparative Models}
\label{}

The proposed model was compared with state-of-the-art models based on ML as well as classical statistical models such as ARIMA and ETS. All ML models except LSTM use a pattern representation of time series \cite{Dud20}. Patterns which express preprocessed repetitive sequences in a time series ensure input and output data unification through trend filtering and variance equalization. Consequently, the relationship between input and output data is simplified and the transformed forecasting problem can be solved using simple models. 

The comparative models are described in detail in \cite{Dud20} and outlined below.

\begin{itemize}
	\item $k$-NNw -- $k$-nearest neighbor weighted regression model. It estimates a vector-valued regression function as an average of output patterns in a varying neighborhood of a query pattern. A weighting function allows the model to take into account the similarity between the query pattern and its nearest neighbors. The model hyperparameters are: input pattern length and number of nearest neighbors $k$. Linear weighting function was used.
	
	\item FNM -- fuzzy neighborhood model. In this case, the regression model is similar to $k$-NNw except that it aggregates not only $k$-nearest neighbors of the query pattern but all training patterns. The weighting function takes the form of a membership function (Gaussian-type function) which assigns to each training pattern a degree of membership to the query pattern neighborhood. The model hyperparameters are: input pattern length and membership function width.
	
	\item N-WE -- Nadaraya–Watson estimator. It is a kernel regression model belonging to the same category of nonparametric models as $k$-NNw and FNM. N-WE estimates the regression function as a locally weighted average, using a kernel function (Gaussian) as a weighting function. The model hyperparameters are: input pattern length and kernel bandwidth parameters.
	
	\item GRNN -- general regression NN model. This is a four-layer NN with Gaussian nodes centered at training patterns.
	The node outputs express similarities between the query pattern and the training patterns. These outputs are treated as the weights of the training patterns in the regression model. The model hyperparameters are: input pattern length and bandwidth parameter for nodes.
	
	\item MLP -- multilayer perceptron \cite{Pel19b} with a single hidden layer and sigmoidal neurons. It used for learning the Levenberg-Marquardt method with Bayesian regularization to prevent overfitting. The MLP hyperparameters are: input pattern length and number of hidden nodes. We use Matlab R2018a implementation of MLP (function \texttt{feedforwardnet} from Neural Network Toolbox).         
	
	\item ANFIS -- adaptive neuro-fuzzy inference system \cite{Pel18}. The initial membership function parameters in the premise parts of rules are determined using fuzzy $c$-means clustering. A hybrid learning method is applied for ANFIS training which uses a combination of the least-squares for consequent parameters and backpropagation gradient descent method for premise parameters. The ANFIS hyperparameters are: input pattern length and number of rules. The Matlab R2018a implementation of ANFIS was used (function \texttt{anfis} from Fuzzy Logic Toolbox).

	\item $k$-NNw+ETS, FNM+ETS, N-WE+ETS, GRNN+ETS, MLP+ETS, ANFIS+ETS -- variants of the above models with output patterns encoded with variables describing the current features of the time series. These features are the mean value of the series in the seasonal cycle and its dispersion. To postprocess the forecasted output pattern, the coding variables are predicted for the next period using ETS. We use R implementation of ETS (see below).
	
	\item LSTM -- long short-term memory, where the responses are the training sequences with values shifted by one time step (a sequence-to-sequence regression LSTM network). For multiple time steps, after one step was predicted the LSTM state was updated. Previous prediction was used as input to LSTM, producing a forecast for the next time step. LSTM was optimized using Adam (adaptive moment estimation) optimizer. The length of the hidden state was the only hyperparameter to be tuned. Other hyperparameters remain at their default values. The experiments were carried out using Matlab R2018a implementation of LSTM (function \texttt{trainNetwork} from Neural Network Toolbox). 
	
	\item ARIMA -- ARIMA$(p, d, q)(P, D, Q)_{12}$ model implemented in function \texttt{auto.arima} in R environment (package \texttt{forecast}). This function implements automatic ARIMA modeling which combines unit root tests, minimization of the Akaike information criterion (AICc) and maximum likelihood estimation to obtain the optimal ARIMA model \cite{Hyn20}.
	
	\item ETS -- exponential smoothing state space model \cite{Hyn08} implemented in function \texttt{ets} (R package \texttt{forecast}). This implementation includes many types of ETS models depending on how the seasonal, trend and error components are taken into account. They can be expressed additively or multiplicatively, and the trend can be damped or not. As in the case of \texttt{auto.arima}, \texttt{ets} returns the optimal model estimating its parameters using AICc.

\end{itemize}

The models: $k$-NNw, FNM, N-WE and GRNN are also known as pattern similarity-based forecasting models (PSFMs) because the forecast is constructed by aggregating the training output patterns using similarity between the query pattern and training input patterns \cite{Dud20}. All the model hyperparameters mentioned above were selected on the training set in grid search procedures. The NN-based models, i.e. MLP, ANFIS and LSTM, due to the stochastic nature of the learning processes return different results for the same data. In this study, 100 independent learning sessions for these models were performed and the final errors were calculated as  averages over 100 trials.

\subsection{Results}

Table \ref{tab1} shows the results of forecasting for the proposed and comparative models: median of absolute percentage error (APE), mean APE (MAPE), interquartile range of APE as a measure of the forecast dispersion, and root mean square error (RMSE). These are values averaged over 35 countries. The lowest errors are for "+ETS" PSBMs, where the coding variables are predicted using ETS. The errors for ETS+RD-LSTM are slightly higher but lower than for all other models.

\begin{table}[]
	\caption{Results comparison among proposed and comparative models.}
	\label{tab1}
	\setlength{\tabcolsep}{9pt}
	\centering
	\begin{tabular}{lcccc}
		\hline
		Model       & \multicolumn{1}{c}{Median \textit{APE}} & \multicolumn{1}{c}{\textit{MAPE}} & \multicolumn{1}{c}{\textit{IQR}} & \multicolumn{1}{c}{\textit{RMSE}} \\ \hline
		k-NNw	&	2.89	&	4.99	&	3.85	&	368.79	\\
		FNM	&	2.88	&	4.88	&	4.26	&	354.33	\\
		N-WE	&	2.84	&	5.00	&	3.97	&	352.01	\\
		GRNN	&	2.87	&	5.01	&	4.02	&	350.61	\\
		k-NNw+ETS	&	2.71	&	4.47	&	3.52	&	327.94	\\
		FNM+ETS	&	2.64	&	4.40	&	3.46	&	321.98	\\
		N-WE+ETS	&	2.68	&	4.37	&	3.36	&	320.51	\\
		GRNN+ETS	&	2.64	&	4.38	&	3.51	&	324.91	\\
		MLP	&	2.97	&	5.27	&	3.84	&	378.81	\\
		MLP+ETS	&	3.11	&	4.80	&	4.12	&	358.07	\\
		ANFIS	&	3.56	&	6.18	&	4.87	&	488.75	\\
		ANFIS+ETS	&	3.54	&	6.32	&	4.26	&	464.29	\\
		LSTM	&	3.73	&	6.11	&	4.50	&	431.83	\\
		ARIMA	&	3.32	&	5.65	&	5.24	&	463.07	\\
		ETS	&	3.50	&	5.05	&	4.80	&	374.52	\\
		ETS+RD-LSTM	&	2.74	&	4.48	&	3.55	&	347.24	\\ 	 \hline
	\end{tabular}
\end{table}

\begin{figure*}[]
	\centering
	\includegraphics[width=0.95\textwidth]{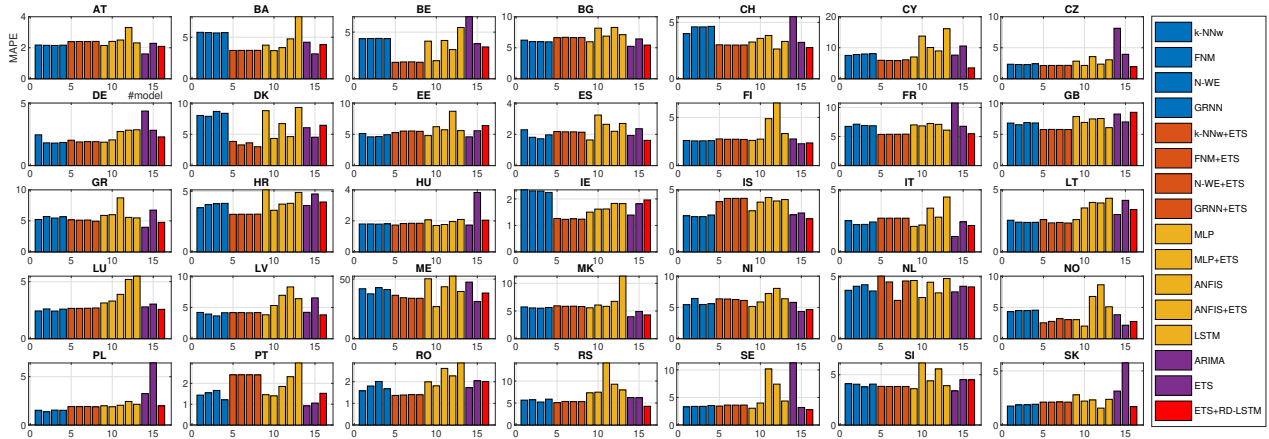}
	\caption{MAPE for each country.} 
	\label{figC}
\end{figure*}

\begin{figure}[]
	\centering
	\includegraphics[width	=0.49\textwidth]{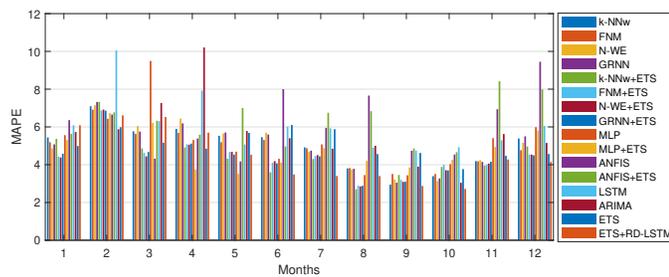}
	\caption{MAPE for each month of the forecasted period.} 
	\label{figM}
\end{figure}

More detailed results are shown in Figs 9 and 10. Fig. \ref{figC} depicts MAPE for each country. As can be seen from this figure, ETS+RD-LSTM is one of the most accurate models in most cases. Fig. \ref{figM} shows MAPE for each month of the forecasted period. Note lower errors for months 8--10 and higher for months 1--4 and 12.  ETS+RD-LSTM achieved better results than most of the comparative models for the months 6--12. For months 1 and 3 it achieved the highest errors compared to other models.  

\begin{figure}[]
	\centering
	\includegraphics[width=0.24\textwidth]{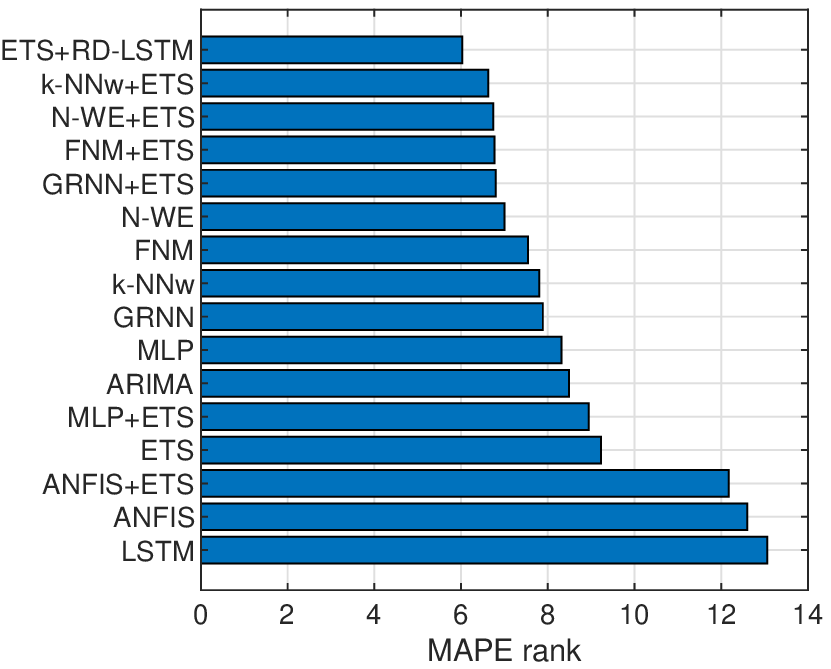}
	\includegraphics[width=0.24\textwidth]{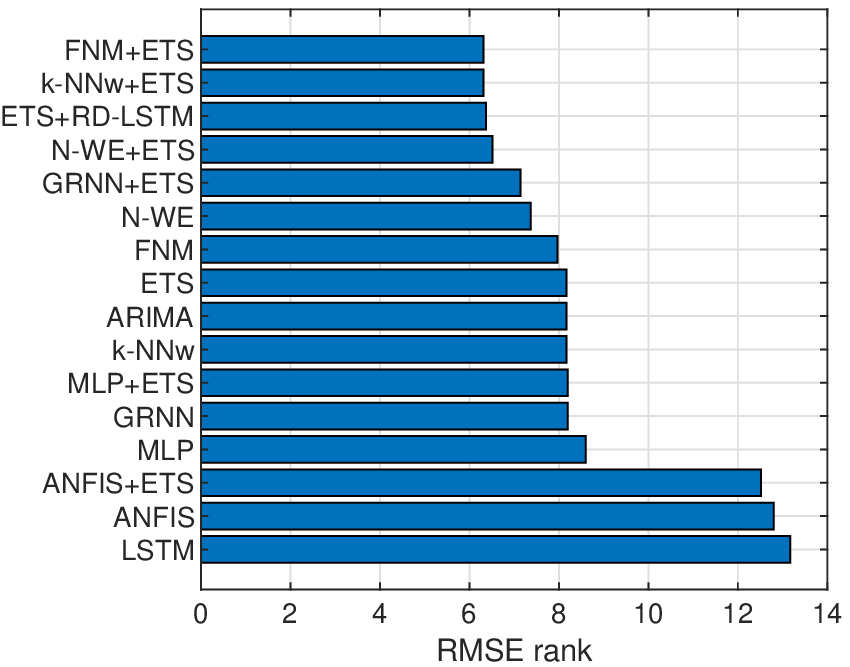}
	\caption{Rankings of the models.} 
	\label{figR}
\end{figure}

Rankings of the models are shown in Fig. \ref{figR}. These are based on average ranks of the models in the rankings for individual countries. The left panel shows the ranking based on MAPE and the right panel shows the ranking based on RMSE. Note the high position of ETS+RD-LSTM. It is in first position in MAPE ranking and third position in RMSE ranking. Note that in the latter case the difference between the first three positions is very small. 

\begin{figure*}[]
	\centering
	\includegraphics[width=0.3\textwidth]{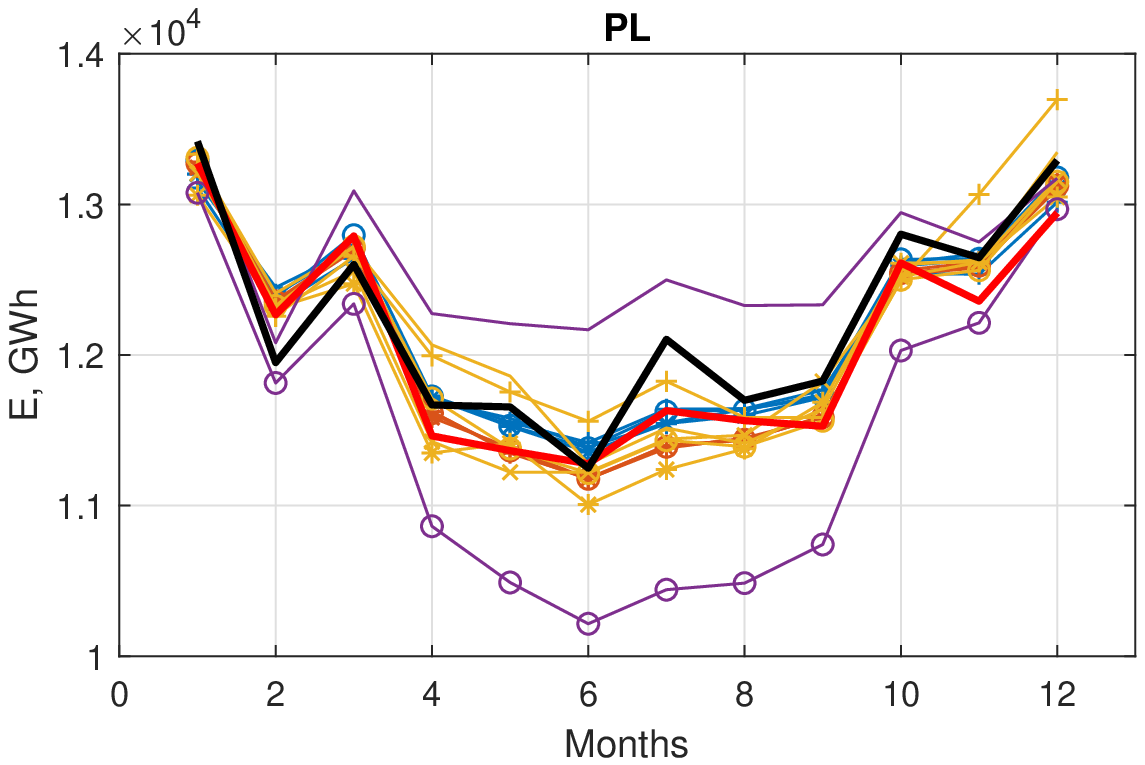}
	\includegraphics[width=0.3\textwidth]{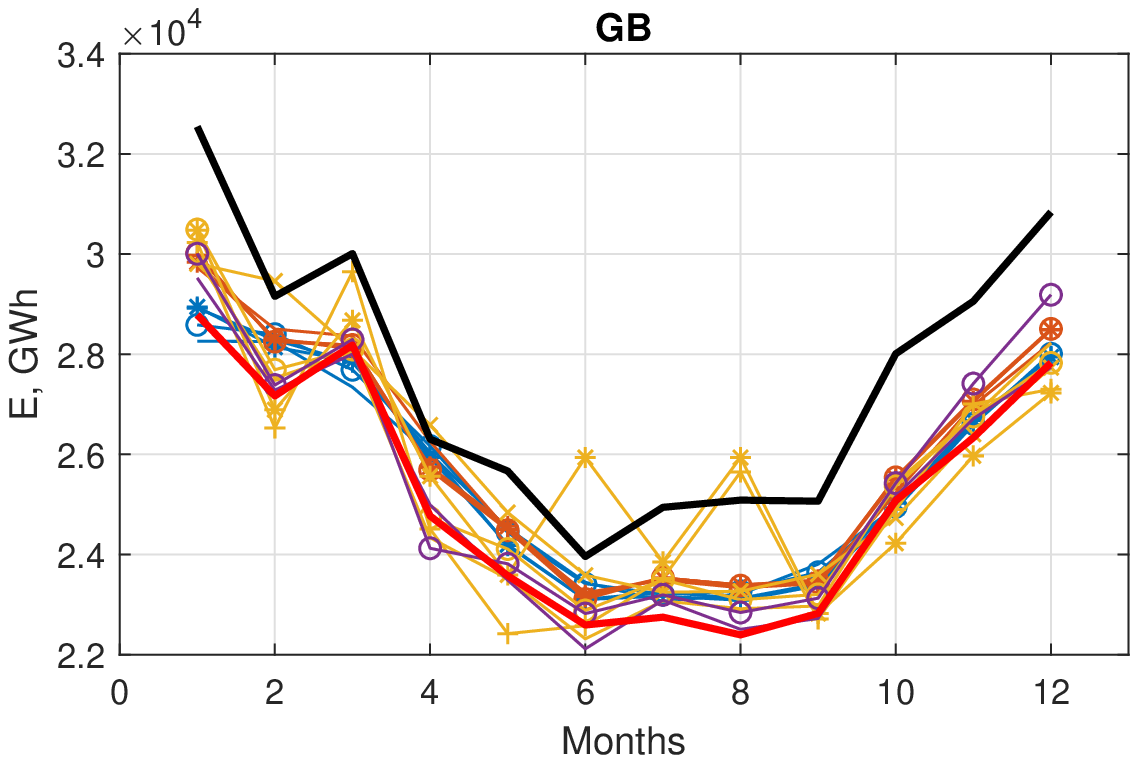}
	\includegraphics[width=0.3\textwidth]{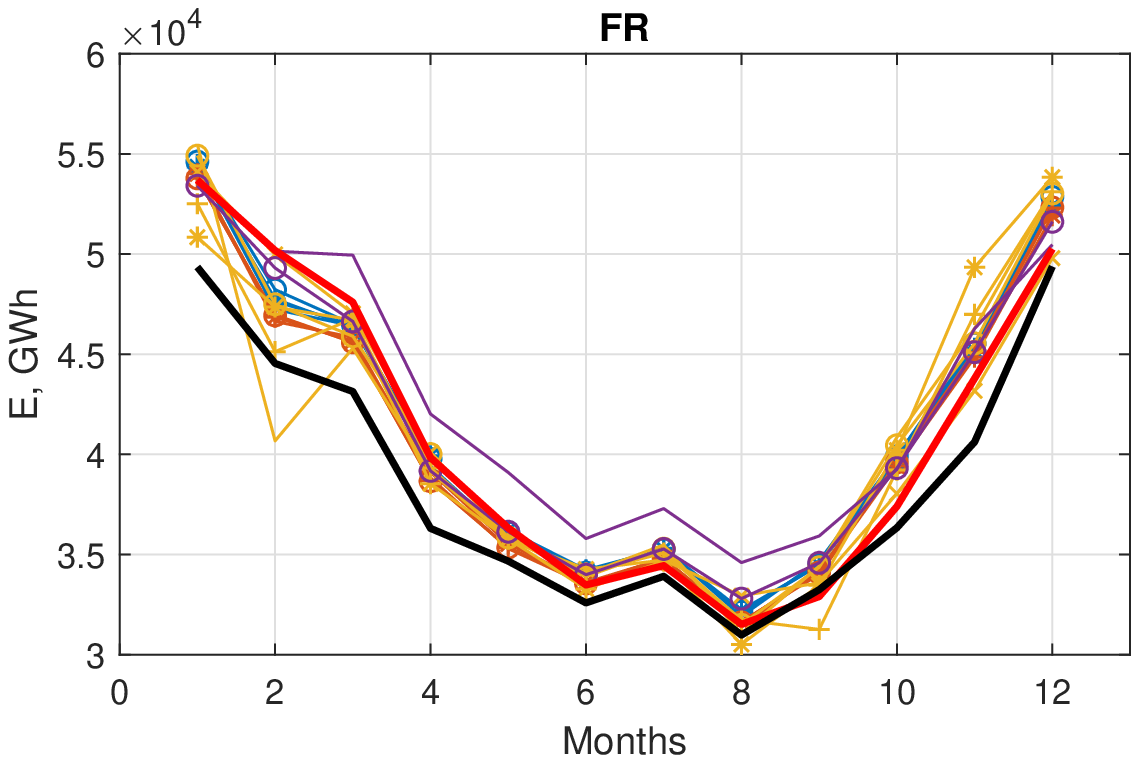}
	\includegraphics[width=0.3\textwidth]{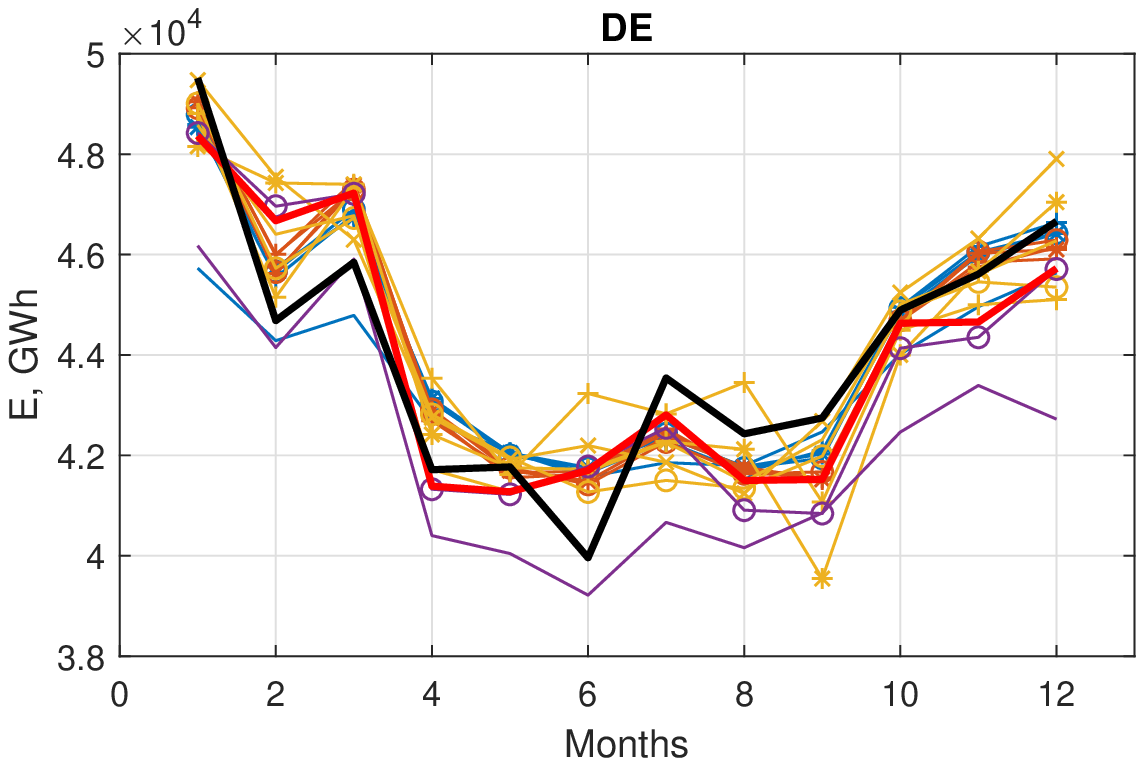}
	\includegraphics[width=0.3\textwidth]{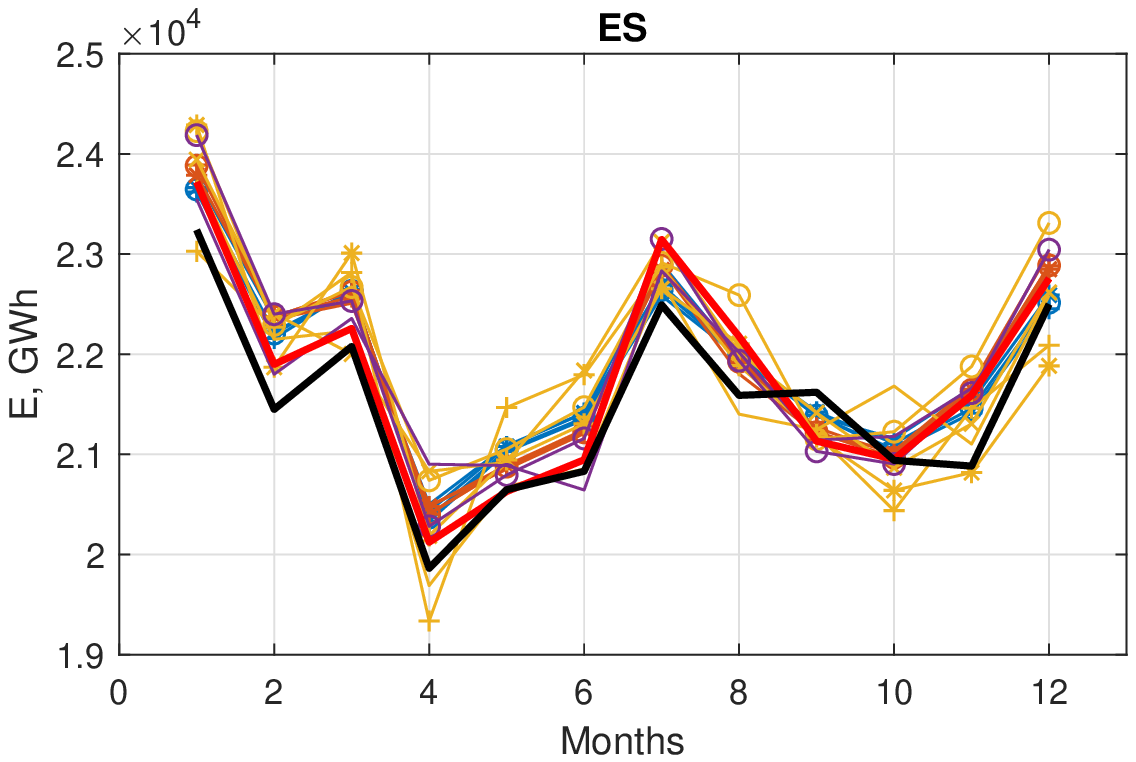}
	\includegraphics[width=0.3\textwidth]{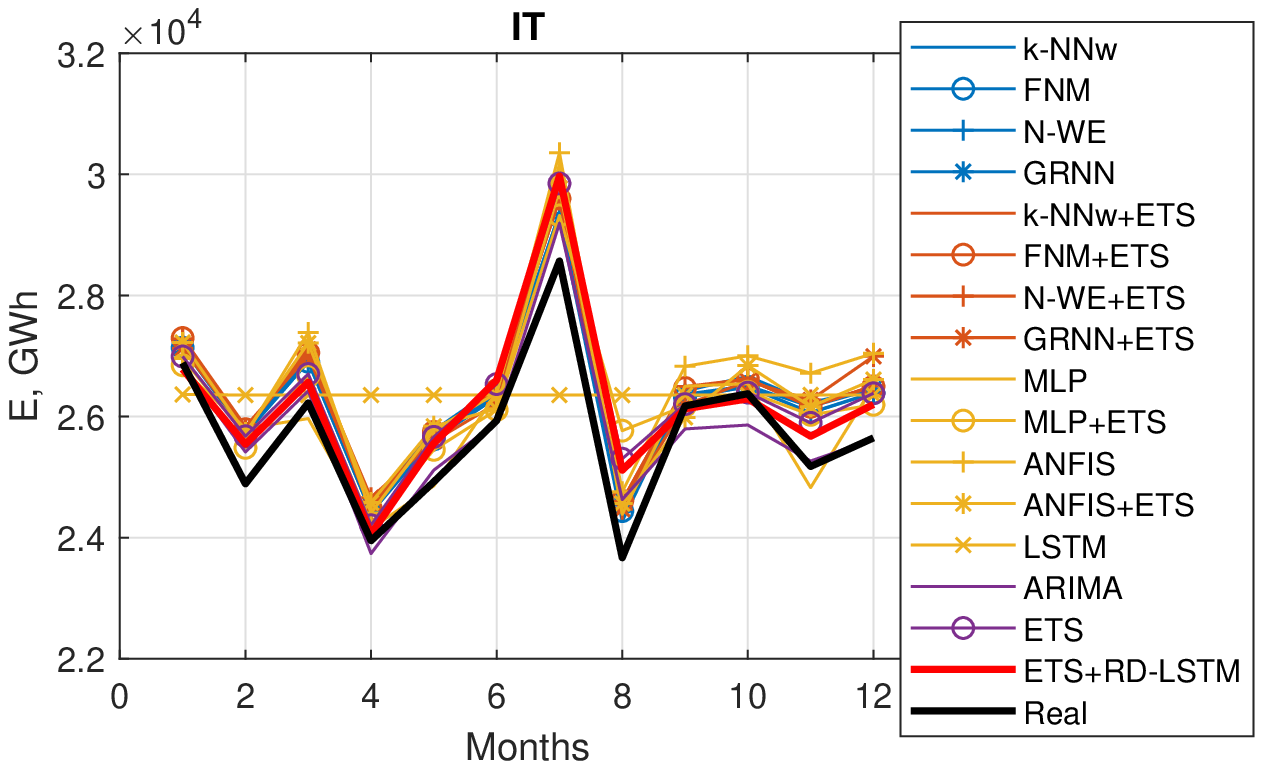}
	\caption{Real and forecasted monthly electricity demand for selected countries.} 
	\label{figF}
\end{figure*}

Examples of forecasts produced by the models for six countries are depicted in Fig. \ref{figF}. For PL data, most models including ETS+RD-LSTM do not exceed $MAPE = 2\%$, which should be considered a very good result. Similar results were achieved for ES, IT and DE. In these cases MAPE for ETS+RD-LSTM were respectively: $1.61\%$ (most accurate model), $2.12\%$ (third most accurate model) and $2.29\%$. 	 
For GB the forecasts are underestimated. This results from
the fact that demand went up unexpectedly in 2014 despite
the downward trend observed in the previous period from
2010 to 2013. The reverse situation for FR caused a slight
overestimation of forecasts. 
For GB data, ETS+RD-LSTM with $MAPE=8.52\%$ was the least accurate model, and for FR data, with $MAPE=5.49\%$, it was one the most accurate models. 

%\begin{figure*}[]
%	\centering
%	\includegraphics[width=0.325\textwidth]{PL.eps}
%	\includegraphics[width=0.325\textwidth]{GB.eps}
%	\includegraphics[width=0.325\textwidth]{FR.eps}
%	\includegraphics[width=0.325\textwidth]{DE.eps}
%	\includegraphics[width=0.325\textwidth]{ES.eps}
%	\includegraphics[width=0.325\textwidth]{IT2.eps}
%	\caption{Real and forecasted monthly electricity demand for selected countries.} 
%	\label{figF}
%\end{figure*}

\begin{figure}[]
	\centering
	\includegraphics[width	=0.4\textwidth]{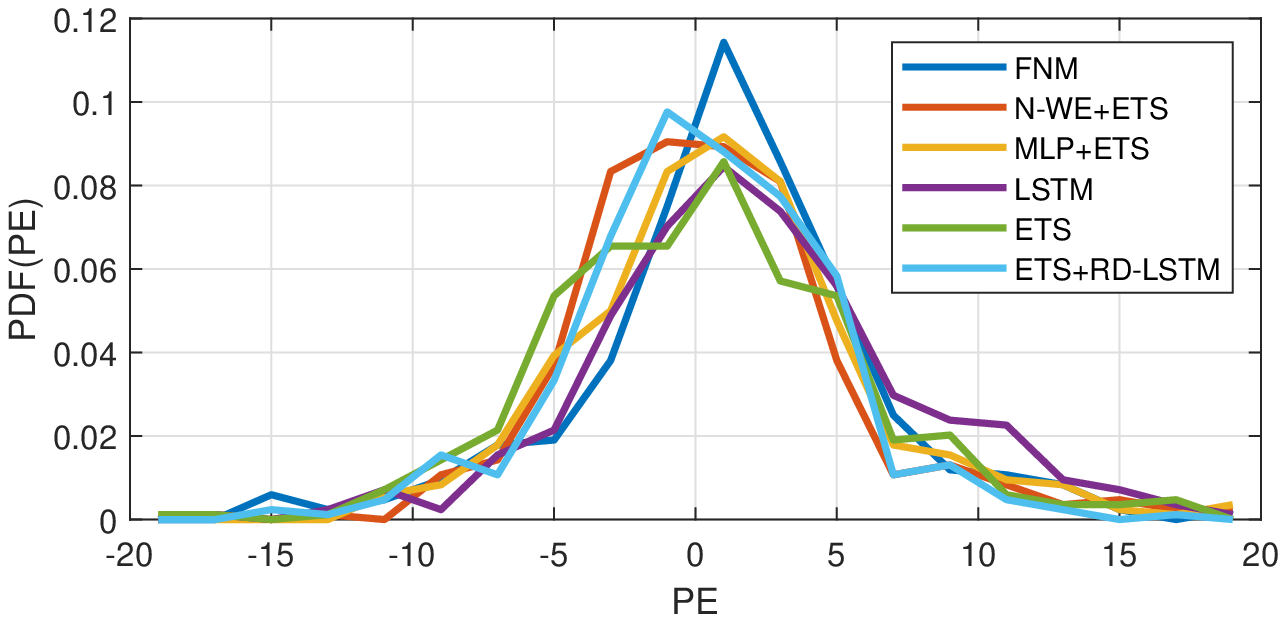}
	\caption{PDF of the percentage errors.} 
	\label{figP}
\end{figure}

Table \ref{tabDS} shows basic descriptive statistics of
the percentage errors (PEs) and Fig. \ref{figP} depicts the empirical probability density functions of PEs for the selected models. 
The PE distributions are similar to normal but tests for the assessment of normality (Jarque–Bera test and
Lilliefors test) do not confirm this. In all cases, the forecasts
are overestimated, having a positive PE mean. Note that ETS+RD-LSTM is one of the least biased models. Positive values
of skewness indicate the right-skewed PE distributions, and
high kurtosis values indicate leptokurtic distributions where
the probability mass is concentrated around the mean.

\begin{table}[]
	\caption{Descriptive statistics of percentage errors.}
	%\begin{center}
	\begin{tabular}{lccccc}
		\hline
		&Mean & Median & Std & Skewness &Kurtosis\\
		\hline
		k-NNw	&	1.87	&	1.08	&	10.43	&	5.14	&	44.56	\\
		FNM	&	2.03	&	1.22	&	9.34	&	4.16	&	35.14	\\
		N-WE	&	1.91	&	1.18	&	10.82	&	5.41	&	48.94	\\
		GRNN	&	1.87	&	1.16	&	10.60	&	5.34	&	48.11	\\
		k-NNw+ETS	&	1.25	&	0.20	&	9.00	&	4.40	&	37.30	\\
		FNM+ETS	&	1.26	&	0.11	&	8.80	&	4.75	&	41.71	\\
		N-WE+ETS	&	1.26	&	0.17	&	8.68	&	4.63	&	40.75	\\
		GRNN+ETS	&	1.26	&	0.11	&	8.61	&	4.42	&	38.38	\\
		MLP	&	1.37	&	0.68	&	11.88	&	7.52	&	109.64	\\
		MLP+ETS	&	1.71	&	1.03	&	7.32	&	1.55	&	11.83	\\
		ANFIS	&	2.51	&	1.43	&	11.37	&	4.35	&	34.93	\\
		ANFIS+ETS	&	1.30	&	0.40	&	12.65	&	0.96	&	39.37	\\
		LSTM	&	3.12	&	1.81	&	9.49	&	2.86	&	22.21	\\
		ARIMA	&	2.35	&	1.03	&	13.62	&	9.01	&	119.20	\\
		ETS	&	1.04	&	0.31	&	7.97	&	1.89	&	13.52	\\
		ETS+RD-LSTM	&	1.11	&	0.27	&	10.07	&	6.37	&	63.61	\\
		\hline
	\end{tabular}
	\label{tabDS}
	%\end{center}
\end{table}
 
\section{Conclusion}

In this work, we proposed and empirically validated a new architecture for mid-term load forecasting. It was inspired by the winning submission to the M4 forecasting competition 2018, which combines ETS, advanced LSTM and ensembling.
The model has a hierarchical structure composed of a global part learned
across many time series (weights of the LSTM) with a time series specific part (ETS smoothing coefficients and initial seasonal components).
Using stochastic gradient descent, it learns a mapping from input vectors to output vectors and time series representations at the same time. 
ETS-inspired formulas, extract the main components of individual time series to be used for deseasonalization and normalization.  
Preprocessed time series are forecasted using residual dilated LSTM. Due to the introduction of dilated recurrent skip connections and a spatial shortcut path from lower layers, LSTM is able to capture better long-term seasonal relationships and ensure more efficient training. To deal with forecast bias we used a pinball loss function with a parameter controlling its asymmetry. In addition, we penalized the loss function to prevent overfitting. Three-level ensembling ensures powerful regularization reducing the model variance, which has sources in the stochastic nature of SGD, and also in data and parameter uncertainty. 

We applied the proposed model to the monthly electricity demand forecasting for 35 European countries. The results demonstrated its state-of-the-art performance and competitiveness with classical models such as ARIMA and ETS as well as state-of-the-art models based on ML.

% or


%\section*{Acknowledgment}
%This work was supported by Grant 2017/27/B/ST6/01804 from the National Science Centre, Poland.

\ifCLASSOPTIONcaptionsoff
  \newpage
\fi


\begin{thebibliography}{10}
\providecommand{\url}[1]{#1}
\csname url@samestyle\endcsname
\providecommand{\newblock}{\relax}
\providecommand{\bibinfo}[2]{#2}
\providecommand{\BIBentrySTDinterwordspacing}{\spaceskip=0pt\relax}
\providecommand{\BIBentryALTinterwordstretchfactor}{4}
\providecommand{\BIBentryALTinterwordspacing}{\spaceskip=\fontdimen2\font plus
\BIBentryALTinterwordstretchfactor\fontdimen3\font minus
  \fontdimen4\font\relax}
\providecommand{\BIBforeignlanguage}[2]{{%
\expandafter\ifx\csname l@#1\endcsname\relax
\typeout{** WARNING: IEEEtran.bst: No hyphenation pattern has been}%
\typeout{** loaded for the language `#1'. Using the pattern for}%
\typeout{** the default language instead.}%
\else
\language=\csname l@#1\endcsname
\fi
#2}}
\providecommand{\BIBdecl}{\relax}
\BIBdecl

\bibitem{Anon19}
G.~Dudek, ``Data-driven randomized learning of feedforward neural networks,''
  \emph{arXiv:1908.03891}, 2019.

\bibitem{Pri15}
J.~Principe and B.~Chen, ``Universal approximation with convex optimization:
  Gimmick or reality?'' \emph{IEEE Comput Intell Mag}, vol.~10, pp. 68--77,
  2015.

\bibitem{Hau95}
D.~Husmeier, ``Random vector functional link (rvfl) networks,'' in \emph{Neural
  Networks for Conditional Probability Estimation: Forecasting Beyond Point
  Predictions}.\hskip 1em plus 0.5em minus 0.4em\relax Springer-Verlag London,
  1999, ch.~6, pp. 87--97.

\bibitem{Zha16c}
L.~Zhang and P.~Suganthan, ``A survey of randomized algorithms for training
  neural networks,'' \emph{Information Sciences}, vol. 364, pp. 146--155, 2016.

\bibitem{Cao18}
W.~Cao, X.~Wang, Z.~Ming, and J.~Gao, ``A review on neural networks with random
  weights,'' \emph{Neurocomputing}, vol. 275, pp. 278--287, 2018.

\bibitem{Li17}
M.~Li and D.~Wang, ``Insights into randomized algorithms for neural networks:
  Practical issues and common pitfalls,'' \emph{Information Sciences}, vol.
  382--383, pp. 170--178, 2017.

\bibitem{Pao94}
Y.~Pao, G.~Park, and D.~Sobajic, ``Learning and generalization characteristics
  of the random vector functional-link net,'' \emph{Neurocomputing}, vol.~6,
  no.~2, pp. 163--180, 1994.

\bibitem{Wan17}
D.~Wang and M.~Li, ``Stochastic configuration networks: Fundamentals and
  algorithms,'' \emph{IEEE Trans. Cybernetics}, vol.~47, no.~10, pp.
  3466--3479, 2017.

\bibitem{Dud19}
G.~Dudek, ``Generating random weights and biases in feedforward neural networks
  with random hidden nodes,'' \emph{Information Sciences}, vol. 481, pp.
  33--56, 2019.

\bibitem{Dud19a}
------, ``Improving randomized learning of feedforward neural networks by
  appropriate generation of random parameters,'' in \emph{Advances in
  Computational Intelligence, IWANN 2019}, I.~Rojas, G.~Joya, and A.~Catala,
  Eds.\hskip 1em plus 0.5em minus 0.4em\relax Cham: Springer International
  Publishing, 2019, pp. 517--530.

\bibitem{Dai19}
W.~Dai, D.~Li, P.~Zhou, and T.~Chai, ``Stochastic configuration networks with
  block increments for data modeling in process industries,'' \emph{Information
  Sciences}, vol. 484, pp. 367--386, 2019.

\end{thebibliography}


\begin{thebibliography}{1}
	
	
\bibitem{Apa12}
F. Apadula, A. Bassini, A. Elli, and S. Scapin, ``Relationships between meteorological variables and monthly electricity demand,`` \textit{Appl Energy}, vol. 98, pp. 346–356, 2012.% doi: 10.1016/j.apenergy.2012.03.053
%F. Apadula, A. Bassini, A. Elli, S. Scapin, Relationships between meteorological variables and monthly electricity demand. Appl Energy 98 (2012) 346–356.% doi: 10.1016/j.apenergy.2012.03.053

\bibitem{Dog16}
E. Dogan, ``Are shocks to electricity consumption transitory or permanent? Sub-national evidence from Turkey,`` \textit{Utilities Policy}, vol. 41 pp. 77–84, 2016. %doi: 10.1016/j.jup.2016.06.007
%E. Dogan, Are shocks to electricity consumption transitory or permanent? Sub-national evidence from Turkey, Utilities Policy 41 (2016) 77–84. %doi: 10.1016/j.jup.2016.06.007

\bibitem{Sug11}
L. Suganthi and A. A. Samuel, ``Energy models for demand forecasting — A review,`` \textit{Renew Sust Energy Rev}, vol. 16(2), pp. 1223–1240, 2002. %doi: 10.1016/j.rser.2011.08.014
%L. Suganthi, A.A. Samuel, Energy models for demand forecasting — A review. Renew Sust Energy Rev 16(2) (2012) 1223–1240. %doi: 10.1016/j.rser.2011.08.014

\bibitem{Bar01}
E. H. Barakat, ``Modeling of nonstationary time-series data. Part II. Dynamic periodic trends,`` \textit{Electr Power Energy Systems}, vol. 23, pp.  63–68, 2001.
%E.H. Barakat, Modeling of nonstationary time-series data. Part II. Dynamic periodic trends. Electr Power Energy Systems 23 (2001) 63–68.

\bibitem{Gon08}
E. González-Romera, M.A. Jaramillo-Morán, and D. Carmona-Fernández, ``Monthly electric energy demand forecasting with neural networks and Fourier series,`` \textit{Energy Conversion and Management}, vol. 49, pp. 3135–3142, 2008.
%E. González-Romera, M.A. Jaramillo-Morán, D. Carmona-Fernández, Monthly electric energy demand forecasting with neural networks and Fourier series, Energy Conversion and Management 49 (2008) 3135–3142.

\bibitem{Chen17}
J. F. Chen, S. K. Lo, and Q. H. Do, ``Forecasting monthly electricity demands: An application of neural networks trained by heuristic algorithms,`` \textit{Information}, vol. 8(1), 31, 2017.
%J.F. Chen, S.K. Lo, Q.H. Do, Forecasting monthly electricity demands: An application of neural networks trained by heuristic algorithms, Information 8(1) (2017) 31.

\bibitem{Gav01}
M. Gavrilas, I. Ciutea, and C. Tanasa, ``Medium-term load forecasting with artificial neural network models,`` Proc. Conf. International Conference and Exhibition on Electricity Distribution, vol. 6, 2001.
%M. Gavrilas, I. Ciutea, C. Tanasa, Medium-term load forecasting with artificial neural network models, IEEE Conf. Elec. Dist. Pub. 6 (2001).

\bibitem{Dov99}
E. Doveh, P. Feigin, and L. Hyams, ``Experience with FNN models for medium term power demand predictions,`` \textit{IEEE Trans. Power Syst.}, vol. 14(2), pp. 538–546, 1999.
%E. Doveh, P. Feigin., L. Hyams, Experience with FNN models for medium term power demand predictions, IEEE Trans. Power Syst. 14 (2) (1999) 538-546.

\bibitem{Pel19}
P. Pe\l ka and G. Dudek, ``Medium-term electric energy demand forecasting using generalized regression neural network,`` Proc. Conf. Information Systems Architecture and Technology ISAT 2018, \textit{Advances in Intelligent Systems and Computing}, vol. 853, pp. 218–227, Springer, Cham, 2019. 
%Pełka P., Dudek G.: Medium-Term Electric Energy Demand Forecasting Using Generalized Regression Neural Network. In: Świątek J., Borzemski L., Wilimowska Z. (eds) Information Systems Architecture and Technology: Proceedings of 39th International Conference on Information Systems Architecture and Technology – ISAT 2018. Advances in Intelligent Systems and Computing, vol 853, pp. 218-227. Springer, Cham 2019.

\bibitem{Ahm19}
T. Ahmad and H. Chen, ``Potential of three variant machine-learning models for forecasting district level medium-term and long-term energy demand in smart grid environment,`` \textit{Energy}, vol. 160, pp. 1008–1020, 2018.
%T. Ahmad, H. Chen, Potential of three variant machine-learning models for forecasting district level medium-term and long-term energy demand in smart grid environment, Energy 160 (2018) 1008-1020.

\bibitem{Pei11}
P. C. Chang, C. Y. Fan, and J. J. Lin, ``Monthly electricity demand forecasting based on a weighted evolving fuzzy neural network approach,`` \textit{Electrical Power and Energy Systems}, vol. 33, pp. 17–27, 2011.
%Chang Pei-Chann, Fan Chin-Yuan, Lin Jyun-Jie, Monthly electricity demand forecasting based on a weighted evolving fuzzy neural network approach,  Electrical Power and Energy Systems 33 (2011), 17–27.

\bibitem{Bae19}
S. Baek, ``Mid-term Load Pattern Forecasting With Recurrent Artificial Neural Network,`` in \textit{IEEE Access}, vol. 7, pp. 172830–172838, 2019.
%S. Baek, "Mid-term Load Pattern Forecasting With Recurrent Artificial Neural Network," in IEEE Access, vol. 7, pp. 172830-172838, 2019.

\bibitem{Zhao12}
W. Zhao, F. Wang, and D. Niu, ``The application of support vector machine in load forecasting,`` \textit{Journal of Computers}, vol. 7(7), pp. 1615–1622, 2012.
%W. Zhao, F. Wang, D. Niu, The application of support vector machine in load forecasting, Journal of Computers 7(7) (2012) 1615-1622.

\bibitem{Dud20}
G. Dudek and P. Pe\l ka, ``Pattern Similarity-based Machine Learning Methods for Mid-term Load Forecasting: A Comparative Study,`` arXiv:2003.01475, 2020.
%G. Dudek, P. Pełka, Pattern Similarity-based Machine Learning Methods for Mid-term Load Forecasting: A Comparative Study. arXiv:2003.01475.


\bibitem{Hew19}
H. Hewamalage, C. Bergmeir, and K. Bandara, ``Recurrent neural networks for time series forecasting: Current status and future directions,`` arXiv:1909.00590v3, 2019.

\bibitem{Yan18}
K. Yan, X. Wang ,Y. Du ,N. Jin , H. Huang, and H. Zhou, ``Multi-Step Short-Term Power Consumption Forecasting with a Hybrid Deep Learning Strategy,`` \textit{Energies}, vol. 11(11), 3089, 2018.% https://doi.org/10.3390/en11113089

\bibitem{Bed18}
J. Bedi and D. Toshniwal, ``Empirical mode decomposition based deep learning for electricity demand forecasting,`` \textit{IEEE Access}, vol. 6, pp. 49144–49156, 2018.

\bibitem{Zhe17}
H. Zheng, J. Yuan, and L. Chen, ``Short-Term Load Forecasting Using EMD-LSTM Neural Networks with a XGBboost Algorithm for Feature Importance Evaluation,`` Energies 2017, vol. 10(8), 1168, 2017.% doi:10.3390/en10081168

\bibitem{Nar17}
A. Narayan and K. W. Hipel, ``Long short term memory networks for short-term electric load forecasting,`` Proc. Conf. IEEE International Conference on Systems, Man, and Cybernetics (SMC), Banff, AB, pp. 2573–2578, 2017.
%doi: 10.1109/SMC.2017.8123012

\bibitem{Tou19}
J. Toubeau, J. Bottieau, F. Vallée, and Z. De Grève, ``Deep learning-based multivariate probabilistic
forecasting for short-term scheduling in power markets,`` \textit{IEEE Transactions on Power Systems}, vol. 34(2), pp. 1203–1215, 2019.
%J. Toubeau, J. Bottieau, F. Vallée, and Z. De Grève. Deep learning-based multivariate probabilistic forecasting for short-term scheduling in power markets. IEEE Transactions on Power Systems, 34(2):1203–1215, March 2019.

\bibitem{Zia18}
T. Zia, and S. Razzaq, ``Residual recurrent highway networks for learning deep sequence
prediction models,`` \textit{Journal of Grid Computing}, Jun 2018.	
%Tehseen Zia and Saad Razzaq. Residual recurrent highway networks for learning deep sequence prediction models. Journal of Grid Computing, Jun 2018.	

\bibitem{Ore20}
B. N. Oreshkin, D. Carpov, N. Chapados, and Y. Bengio, ``N-BEATS: Neural basis expansion analysis for interpretable time series forecasting,`` arXiv:1905.10437v4, 2020.
%Boris N. Oreshkin, Dmitri Carpov, Nicolas Chapados, Yoshua Bengio, N-BEATS: Neural basis expansion analysis for interpretable time series forecasting. arXiv:1905.10437v4

\bibitem{Sal19}
D. Salinas, V. Flunkert, and J. Gasthaus,
``DeepAR: Probabilistic Forecasting with Autoregressive Recurrent Networks,``
arXiv:1704.04110v3, 2019.
%David Salinas, Valentin Flunkert, Jan Gasthaus DeepAR: Probabilistic Forecasting with Autoregressive Recurrent Networks arXiv:1704.04110v3

\bibitem{Ran18}
S. S. Rangapuram, M. W. Seeger, J. Gasthaus, L. Stella, Y. Wang, and
T. Januschowski, ``Deep state space models for time series forecasting`` In \textit{NeurIPS}, vol. 31, pp.
7785–7794, 2018.
%Syama Sundar Rangapuram, Matthias W Seeger, Jan Gasthaus, Lorenzo Stella, Yuyang Wang, and Tim Januschowski. Deep state space models for time series forecasting. In NeurIPS 31, pp.7785–7794, 2018.

\bibitem{Mar18a}
S. Makridakis, E. Spiliotis, and V.Assimakopoulos, ``The M4 Competition: Results, findings, conclusion and way forward,`` \textit{International Journal of Forecasting}, vol. 34(4). pp. 802–808, 2018. %doi:10.1016/j.ijforecast.2018.06.001.
%Makridakis, Spyros; Spiliotis, Evangelos; Assimakopoulos, Vassilios (October 2018). "The M4 Competition: Results, findings, conclusion and way forward". International Journal of Forecasting. 34 (4): 802–808. %doi:10.1016/j.ijforecast.2018.06.001.

\bibitem{Mar20}
S. Makridakis, E. Spiliotis, and V.Assimakopoulos, ``The M4 Competition: 100,000 time series and 61 forecasting methods,`` \textit{International Journal of Forecasting}, vol. 36(1): pp. 54–74, 2020.
%Makridakis, Spyros; Spiliotis, Evangelos; Assimakopoulos, Vassilios (January 2020). "The M4 Competition: 100,000 time series and 61 forecasting methods". International Journal of Forecasting. 36 (1): 54–74.

\bibitem{Smy20}
S. Smyl, ``A hybrid method of exponential smoothing and recurrent neural networks for time series forecasting,``
\textit{International Journal of Forecasting}, vol. 36(1), pp. 75–85, 2020.
%Slawek Smyl,A hybrid method of exponential smoothing and recurrent neural networks for time series forecasting,International Journal of Forecasting,Volume 36, Issue 1,2020, Pages 75-85,

\bibitem{Cha17}
S. Chang, Y. Zhang, W. Han., M. Yu, X. Guo, W. Tan, et al.
``Dilated recurrent neural networks,`` arXiv:
1710.02224, 2017.
%Chang, S., Zhang, Y., Han, W., Yu, M., Guo, X., Tan, W., et al.(2017). Dilated recurrent neural networks. arXiv e-prints, arXiv:1710.02224.
\bibitem{Kim17}
J. Kim, M. El-Khamy, and J. Lee, ``Residual LSTM: Design of a deep
recurrent architecture for distant speech recognition, ``
arXiv:1701.03360, 2017.
%Kim, J., El-Khamy, M., & Lee, J. (2017). Residual LSTM: Design of a deeprecurrent architecture for distant speech recognition. arXiv e-prints,arXiv:1701.03360.

\bibitem{Qin17}
Y. Qin, D. Song, H. Chen, W. Cheng, G. Jiang, and G. W. Cottrell. ``A dual-stage attention-based recurrent neural network for time series prediction,`` In \textit{IJCAI-17}, pp.
2627–2633, 2017.
%Yao Qin, Dongjin Song, Haifeng Chen, Wei Cheng, Guofei Jiang, and Garrison W. Cottrell. A dual-stage attention-based recurrent neural network for time series prediction. In IJCAI-17, pp.2627–2633, 2017.
\bibitem{Hyn20}
R. J. Hyndman and G. Athanasopoulos, Forecasting: principles and practice, 2nd edition, OTexts: Melbourne, Australia. OTexts.com/fpp2 (2018) Accessed on 7 March 2020.
%Hyndman, R. J., Athanasopoulos, G.: Forecasting: principles and practice. 2nd edition, OTexts: Melbourne, Australia (2018). OTexts.com/fpp2  Accessed on 7 March 2020.

\bibitem{Dud15a}
G. Dudek, ``Pattern Similarity-based Methods for Short-term Load Forecasting – Part 1: Principles,`` \textit{Applied Soft Computing}, vol. 37, pp. 277–287, 2015.
%Dudek G.: Pattern Similarity-based Methods for Short-term Load Forecasting – Part 2: Models. Applied Soft Computing, vol. 36, pp. 422-441, 2015.

\bibitem{Dud15}
G. Dudek, ``Pattern Similarity-based Methods for Short-term Load Forecasting – Part 1: Principles,`` \textit{Applied Soft Computing}, vol. 37, pp. 277–287, 2015.
%Dudek G.: Pattern Similarity-based Methods for Short-term Load Forecasting – Part 1: Principles. Applied Soft Computing, vol. 37, pp. 277-287, 2015.

\bibitem{Hoch97}
S. Hochreiter and J. Schmidhuber, ``Long short-term memory,`` \textit{Neural Computation}, vol. 9(8), pp.1735–1780, 1997.

\bibitem{Gre17}
K. Greff, R. K. Srivastava, J. Koutník, B. R. Steunebrink and J. Schmidhuber, ``LSTM: A Search Space Odyssey,`` in \textit{IEEE Transactions on Neural Networks and Learning Systems}, vol. 28, no. 10, pp. 2222–2232, Oct. 2017.

\bibitem{Pet18}
F. Petropoulos, R. J. Hyndman, and C. Bergmeir, ``Exploring
the sources of uncertainty: Why does bagging for time series
forecasting work?,`` \textit{European Journal of Operational Research}, vol. 268(2),
pp. 545–554, 2008.
%Petropoulos, F., Hyndman, R. J., Bergmeir, C.: Exploring the sources of uncertainty: Why does bagging for time series forecasting work?. European Journal of Operational Research, \textbf{268}(2), 545–554 (2018)

\bibitem{Cha18}
F. Chan and L. L. Pauwels, ``Some theoretical results on forecast
combinations,`` \textit{International Journal of Forecasting}, vol. 34(1), pp. 64–74, 2018.
%Chan, F., Pauwels, L. L.: Some theoretical results on forecast combinations. International Journal of Forecasting, \textbf{34}(1), 64--74 (2018)

\bibitem{Tak06}
I. Takeuchi, Q. V. Le, T. D. Sears, and A. J. Smola, ``Nonparametric quantile estimation`` \textit{Journal of Machine Learning Research}, vol. 7, pp. 1231–1264, 2006.
%Takeuchi, I., Le, Q. V., Sears, T. D., & Smola, A. J. (2006). Nonparametric quantile estimation. Journal of Machine Learning Research (JMLR), 7, 1231–1264.

\bibitem{Neu17}
G. Neubig, C. Dyer, Y. Goldberg, A. Matthews, W. Ammar, A. Anastasopoulos, et al., ``DyNet: The dynamic neural network
toolkit,`` arXiv:1701.03980, 2017
%Neubig, G., Dyer, C., Goldberg, Y., Matthews, A., Ammar, W., Anastasopoulos, A., et al. (2017). DyNet: The dynamic neural network toolkit. arXiv preprint, arXiv:1701.03980.

\bibitem{Pel19b}
P. Pe\l ka and G. Dudek, ``Pattern-based forecasting monthly electricity demand using multilayer perceptron,`` Proc. Conf. Artificial Intelligence and Soft Computing ICAISC 2019,  \textit{Lecture Notes in Artificial Intelligence}, vol. 11508, pp. 663–672, Springer, Cham, 2019. 
%Pełka P., Dudek G.: Pattern-Based Forecasting Monthly Electricity Demand Using Multilayer Perceptron. In: Rutkowski L., Scherer R., Korytkowski M., Pedrycz W., Tadeusiewicz R., Zurada J. (eds) Artificial Intelligence and Soft Computing. ICAISC 2019. LNAI 11508, pp. 663-672. Springer, Cham 2019. 

\bibitem{Pel18}
P. Pe\l ka and G. Dudek, ``Neuro-Fuzzy System for Medium-term Electric Energy Demand Forecasting,`` Proc. Conf. Information Systems Architecture and Technology ISAT 2017, \textit{ Advances in Intelligent Systems and Computing}, vol. 655, pp. 38–47, Springer, Cham, 2018.
%P. Pełka, and G. Dudek, ``Neuro-Fuzzy System for Medium-term Electric Energy Demand Forecasting.,` In: Borzemski L., Świątek J., Wilimowska Z. (eds) Information Systems Architecture and Technology: Proceedings of 38th International Conference on Information Systems Architecture and Technology – ISAT 2017. Advances in Intelligent Systems and Computing, vol 655, pp. 38-47. Springer, Cham, 2018.


\bibitem{Mar18}
S. Makridakis, E. Spiliotis, and V. Assimakopoulos, ``Statistical and machine learning forecasting methods: Concerns and ways forward,`` \textit{PLoS ONE}, vol. 13(3), 2018.
%S Makridakis, E Spiliotis, and V Assimakopoulos. Statistical and machine learning forecasting methods: Concerns and ways forward. PLoS ONE, 13(3), 2018a.

\bibitem{Hyn08}
R. J. Hyndman, A. B. Koehler, J. K. Ord, R. D. Snyder, Forecasting with exponential smoothing: The state space approach, Springer, 2008.
%R.J. Hyndman, A.B. Koehler, J.K. Ord, R.D. Snyder, Forecasting with exponential smoothing: The state space approach, Springer, 2008.	
	

%CNN - https://reader.elsevier.com/reader/sd/pii/S037877961930344X?token=1E256BA0CA2579B5481D638995E7315CB6E1E7AC4CAE3AB77DE76F6B5D41C038F33683F50C09BFA6617F112012686465
	
	
	


\end{thebibliography}
\end{document}